\documentclass[a4paper,preprint]{aastex}

\usepackage{natbib}
\def\blender{{\tt BLENDER}}
\def\kepler{{\slshape Kepler}}
\def\spitzer{{\it Spitzer}}

\begin{document}

\title{The Kepler-19 System: A Transiting 2.2 $R_{\oplus}$ Planet and a Second Planet Detected via Transit Timing Variations}

\author{Sarah~Ballard\altaffilmark{1}, Daniel~Fabrycky\altaffilmark{2}, Francois Fressin\altaffilmark{1}, David~Charbonneau\altaffilmark{1}, Jean-Michel~Desert\altaffilmark{1}, Guillermo~Torres\altaffilmark{1}, Geoffrey Marcy\altaffilmark{3}, Christopher~J.~Burke\altaffilmark{4}, Howard~Isaacson\altaffilmark{3}, Christopher~Henze\altaffilmark{4}, Jason~H.~Steffen\altaffilmark{5},  David~R.~Ciardi\altaffilmark{6}, Steven~B.~Howell\altaffilmark{7,4}, William~D.~Cochran\altaffilmark{8}, Michael~Endl\altaffilmark{8}, Stephen~T.~Bryson\altaffilmark{4},  Jason~F.~Rowe\altaffilmark{4},  Matthew~J.~Holman\altaffilmark{1}, Jack~J.~Lissauer\altaffilmark{4}, Jon~M.~Jenkins\altaffilmark{9}, Martin~Still\altaffilmark{4}, Eric~B.~Ford\altaffilmark{10}, Jessie~L.~Christiansen\altaffilmark{4}, Christopher~K.~Middour\altaffilmark{4}, Michael~R.~Haas\altaffilmark{4}, Jie~Li\altaffilmark{9}, Jennifer~R.~Hall\altaffilmark{11}, Sean~McCauliff\altaffilmark{11}, Natalie~M.~Batalha\altaffilmark{12}, David~G.~Koch\altaffilmark{4}, William~J.~Borucki\altaffilmark{4}}

\altaffiltext{1}{Harvard-Smithsonian Center for Astrophysics, 60 Garden Street, Cambridge, MA 02138, USA; sballard@cfa.harvard.edu}
\altaffiltext{2}{Department of Astronomy and Astrophysics, University of California, Santa Cruz, Santa Cruz, CA 95064, USA}
\altaffiltext{3}{University of California, Berkeley, Berkeley, CA 94720, USA}
\altaffiltext{4}{NASA Ames Research Center, Moffett Field, CA 94035, USA}
\altaffiltext{5}{Fermilab Center for Particle Astrophysics, P.O. Box 500, Batavia, IL 60510, USA}
\altaffiltext{6}{NASA Exoplanet Science Institute/Caltech, Pasadena, CA 91125}
\altaffiltext{7}{National Optical Astronomy Observatory, 950 N. Cherry Ave, Tucson, AZ 85719}
\altaffiltext{8}{McDonald Observatory, The University of Texas, Austin, TX 78712}
\altaffiltext{9}{SETI Institute/NASA Ames Research Center, Moffett Field, CA 94035, USA}
\altaffiltext{10}{Astronomy Department, University of Florida, Gainesville, FL 32111, USA}
\altaffiltext{11}{Orbital Sciences Corporation/NASA Ames Research Center, Moffett Field, CA 94035, USA}
\altaffiltext{12}{San Jose State University, San Jose, CA 95192, USA}


\begin{abstract}
We present the discovery of the Kepler-19 planetary system, which we first identified from a 9.3-day periodic transit signal in the \kepler\ photometry. From high-resolution spectroscopy of the star, we find a stellar effective temperature $T_{\rm eff}$=5541 $\pm$ 60 K, a metallicity [Fe/H]=$-0.13\pm0.06$, and a surface gravity log(g)=$4.59\pm0.10$. We combine the estimate of $T_{\rm eff}$ and [Fe/H] with an estimate of the stellar density derived from the photometric light curve to deduce a stellar mass of $M_{\star}=0.936\pm0.040$ $M_{\odot}$ and a stellar radius of $R_{\star}=0.850\pm0.018$ $R_{\odot}$ (these errors do not include uncertainties in the stellar models). We rule out the possibility that the transits result from an astrophysical false positive by first identifying the subset of stellar blends that reproduce the precise shape of the light curve. Using the additional constraints from the measured color of the system, the absence of a secondary source in the high-resolution spectrum, and the absence of a secondary source in the adaptive optics imaging, we conclude that the planetary scenario is more than three orders of magnitude more likely than a blend. The blend scenario is independently disfavored by the achromaticity of the transit: we measure a transit depth with \spitzer\ at 4.5 $\mu$m of $547^{+113}_{-110}$ ppm, consistent with the depth measured in the \kepler\ optical bandpass of 567$\pm$6 ppm (corrected for stellar limb-darkening). We determine a physical radius of the planet Kepler-19b of $R_{p}=2.209\pm0.048$ $R_{\oplus}$; the uncertainty is dominated by uncertainty in the stellar parameters. From radial-velocity observations of the star, we find an upper limit on the planet mass of 20.3 $M_{\oplus}$, corresponding to a maximum density of 10.4 g cm$^{-3}$. We report a significant sinusoidal deviation of the transit times from a predicted linear ephemeris, which we conclude is due to an additional perturbing body in the system. We cannot uniquely determine the orbital parameters of the perturber, as various dynamical mechanisms match the amplitude, period, and shape of the transit timing signal and satisfy the host star's radial velocity limits.  However, the perturber in these mechanisms has period $\lesssim160$~days and mass $\lesssim6 M_{Jup}$, confirming its planetary nature as Kepler-19c.  We place limits on the presence of transits of Kepler-19c in the available \kepler\ data.
\end{abstract}


\section{Introduction}
With the recent discoveries of the first transiting exoplanets intermediate in size between Earth and Neptune, namely CoRoT-7b \citep{Leger09}, GJ~1214b \citep{Charbonneau09}, Kepler-9d \citep{Torres11}, Kepler-10bc \citep{Batalha11,Fressin11}, Kepler-11bcdf \citep{Lissauer11}, and 55 Cancri b \citep{Winn11}, astronomers have begun in earnest to probe this radius regime of exoplanets, for which no Solar System analog exists. \cite{Borucki11} presents a catalog of 1235 transiting planetary candidates, of which nearly 300 have a radius estimate in the range 1.25$<R_{p}<$2.0 $R_{\oplus}$. While most of these candidates have not yet been confirmed as authentic planets, \cite{Morton11} have shown that the rate of false positives is expected to be low for the \kepler-identified sample. The composition of such planets may be widely variable, as exemplified by the case of GJ~1214b \citep{Charbonneau09}, for which the measured radius and mass were consistent with both a hydrogen envelope or a pure CO$_{2}$ or H$_{2}$O atmosphere \citep{Millerricci10}. It was only with follow-up studies of the atmosphere in transmission that it became possible to distinguish among the various possibilities for the composition of GJ~1214b \citep{Bean10,Croll11,Desert11}. 

The limiting precision of the current state-of-the-art radial velocity observations (meters per second) presents a challenge for the dynamical confirmation of these small planets. In the case of the 1.42 $R_{\oplus}$ transiting planet Kepler-10b, \cite{Batalha11} gathered 40 high-resolution spectra at the Keck telescope \citep{Vogt94} to determine a mass of 4.56$^{+1.17}_{-1.29}$ $M_{\oplus}$. In the absence of radial velocity confirmation, however, it is still possible to make a statistical argument for the planetary nature of the candidate, if the combined likelihood of all false positive scenarios (namely, blends of stars containing an eclipsing member) is sufficiently smaller than the planet scenario. This process of ``validation'' for \kepler-identified planetary candidates has already been applied to three planets in the 1.5-3 $R_{\oplus}$ radius range: Kepler-9d \citep{Torres11}, Kepler-11f \citep{Lissauer11}, and Kepler-10c \citep{Fressin11}. 

Transiting planets are also of interest as they present an opportunity to identify yet more planets in the system by the method of transit timing variations (TTVs). Since this method was proposed \citep{Holman05,Agol05}, the search for planets by TTVs has been a major activity in exoplanet research. \cite{Steffen05} applied a lack of significant TTV variations in the TrES-1 system to deduce constraints on the existence of additional, non-transiting planets. Subsequent works, such as \cite{2008MR} for HD~189733, using $MOST$ observations, \cite{2009B} for CoRoT-1, using the CoRoT satellite, \cite{2009G} for TrES-3, using observations gathered at the Liverpool Telescope, and \cite{Ballard10} for GJ~436, using $EPOXI$ observations, have also used transit times to rule out companions, specifically companions in resonances, for which the TTV method is particularly sensitive to low-mass planets. The \kepler\ team has presented two cases of transit timing variations in exoplanetary systems: Kepler-9  \citep{Holman10} and Kepler-11 \citep{Lissauer11}. In both of these cases, the additional planet (or planets) responsible for the transit timing signal also transit, which enabled mutual constraints on the masses of the planets as predicted by \cite{Holman05}. Over the past year, the \kepler\ team has also presented instances of single-transiting candidate systems showing transit timing variations, but has not confirmed the planetary nature of the candidates or perturbers \citep{Ford11}. Meanwhile, using ground-based observations, several groups have described their transit times as being inconsistent with a linear ephemeris, though two such claims have been revisited by groups who could not confirm the result. In the case of HAT-P-13, while \cite{Pal11} and \cite{Nascimbeni11} found evidence for a companion from the transit times of HAT-P-13b, \cite{Fulton11} demonstrated that the times were consistent with a linear ephemeris (with the exception of a single transit). In the case of OGLE-111b, \cite{Diaz08} claimed that the transit times were inconsistent with a linear ephemeris, but an analysis by \cite{Adams10} with additional transit observations found no evidence for TTVs or duration variations and pointed to systematic errors in previous photometry. \cite{Maciejewski10} and \cite{Fukui11} presented evidence for transit timing variations in the WASP-3 and WASP-5 systems, respectively, but cautioned that additional transits are necessary to confirm or refute the signal (\citealt{Fukui11} expressed caution about unknown systematic effects). \cite{Maciejewski11} presented evidence for TTVs of WASP-10b, and they reported a two-planet orbital solution that fit the TTVs and radial velocities better than alternative orbital models they found, which was not achieved in prior work.

In this paper, we present the discovery of two planets orbiting Kepler-19. The star, which has right ascension and declination 19h21m40.99s and +37d51m06.5s, \kepler\ magnitude $Kp$=11.90, and \kepler\ Input Catalog number 2571238, was identified to host a planetary candidate in the catalog of 1235 \kepler\ identified candidates published by \cite{Borucki11}. In that work, the star was identified by the \kepler\ Object of Interest (KOI) designation KOI-84. The first planet, identified by its transits, has a period of 9.3 days and a radius of 2.2 $R_{\oplus}$, as we discuss below. We validate the planetary nature of the transit signal by a blend analysis. The second planet, Kepler-19c, was identified by transit timing variations. We see no evidence for transits of Kepler-19c in the available \kepler\ data. This detection differs from the ones using ground-based data, summarized in the previous paragraph, in several ways. In the case of Kepler-19, the transiting object has a radius of only 2.2 $R_{\oplus}$, whereas other claims are for perturbations to the transit times of hot Jupiters. Additionally, as we show below, we have well-sampled the TTV signal, since we have measured the transit times of Kepler-19b at a cadence 30 times shorter than the TTV period. Finally, the TTV signal reported here is a much higher signal-to-noise detection. In Section 2, we present the \kepler\ time series from which we detected the system. In Section 3, we present our characterization of the Kepler-19 system from the photometry. And in Section 4, we summarize our follow-up observations of the star. In Section 5, we present the validation of Kepler-19b as a planet. In Section 6, we discuss our constraints on the nature of the perturbing planet Kepler-19c from transit timing variations, as well as our inferred constraints on the composition of the transiting planet Kepler-19b. 

\section{{\slshape Kepler} Observations}
The {\slshape Kepler} spacecraft, launched on 7 March 2009, will photometrically monitor 170,000 stars for 3.5 years for evidence of transiting planets. \cite{VanCleve09} and \cite{Argabright08} give an overview of the {\slshape Kepler} instrument, and \cite{Caldwell10} and \cite{Jenkins10} provide a summary of its performance since launch. The {\slshape Kepler} observations of Kepler-19 that we present in this work were gathered from 5 May 2009 to 5 March 2011. This range spans \kepler\ ``Quarters'' 0--8; \kepler\ operations are divided into four quarters each year. At the end of each quarter, \kepler\ rotates the spacecraft by 90$^{\circ}$ to maintain illumination of the solar panels (though the Quarter 0 pre-science commissioning data were gathered in the same configuration as Quarter 1).  For Quarters 3--8, the observations of Kepler-19 were gathered continuously with an exposure time of 58.8 seconds, corresponding to the ``short cadence'' (SC) mode of the instrument, described by \cite{Jenkins10} and \cite{Gilliland10}, while data from Quarters 0, 1, and 2 were gathered in long-cadence mode (characterized by an exposure time of 29.5 minutes). The data contain gaps of approximately 3 days between quarters for scheduled downlinks. We used the raw light curves generated by the \kepler\ aperture photometry (PA) pipeline, described in \cite{Twicken10}, to which we add two additional steps. First, we remove the effects of baseline drift by individually normalizing each transit. We fit a line with time to the flux immediately before and after transit (specifically, from 9 hours to 20 minutes before first contact, equal to 2.5 transit durations, and an equivalent time after fourth contact). We divide this line from the observations spanning five transit durations and centered on the predicted transit time. For the observations gathered outside of transit, we apply a median filter, with width equal to one day, in order to remove baseline drift over timescales of days. We observed slight flux offsets between observations that occurred after gaps of larger than 1 hour (either for data download, quarterly rolls, or safe modes). The \kepler\ time series of Kepler-19 are shown in Figure 1. 

\begin{figure}[h!]
\begin{center}
 \includegraphics[width=4in]{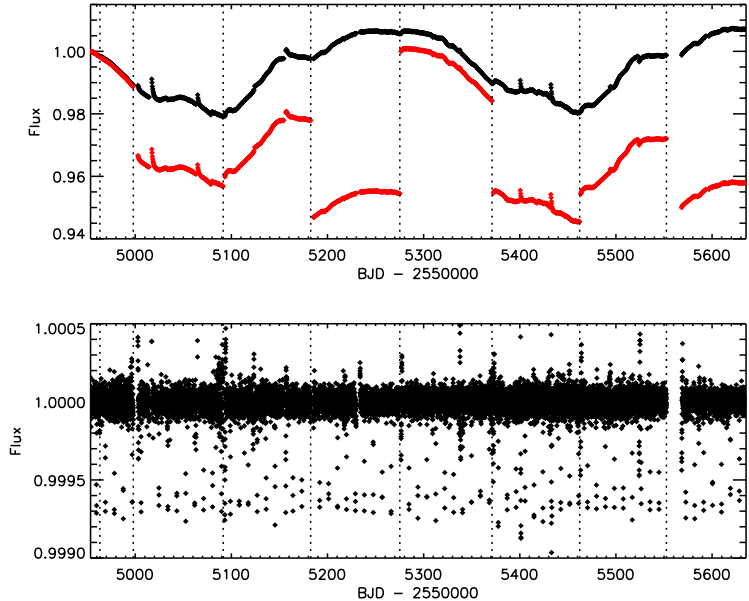} 
 \caption{Kepler-19 relative photometry from Quarters 0-8, as a function of barycentric Julian Day. Data from quarters 0, 1, and 2 were gathered in long-cadence observing mode, while data from quarters 3-8 were gathered in the short-cadence mode. {\it Top panel:} The raw flux depicted in red (with bin size of 1/3 days). We have multiplied the short-cadence observations by a factor of 30 to account for the exposure time ratio between modes, so that all observations appear on the same scale. The gaps between quarters are depicted by dashed lines. To remove the flux offsets between quarterly rolls, we compare the mean brightness during the first two hours of each quarter to the mean of final two hours of the previous quarter, and divide this ratio value from the flux over the entire quarter. This corrected flux is shown in black. {\it Bottom panel:} the detrended \kepler\ light curve, after applying a median filter with width equal to one day and normalizing individual transits as described above. This light curve is depicted with a binsize of 1 hour, so that individual transit events are apparent. Gaps of longer than one hour in the observations are associated with instances of correlated noise in the corrected light curve.}
  \label{fig:alldata}
\end{center}
\end{figure}

\section{Analysis}

\subsection{Derivation of Planetary Parameters from the \kepler\ Light Curve}

 The traditional procedure for fitting the phased transit light curve relies on the assumption of a linear ephemeris. In this case, that assumption does not hold: the transit times deviate from a linear ephemeris in a nearly sinusoidal manner, with a period of about 316 days and an amplitude of about 5 minutes (as compared to a transit duration of 3.5 hours). We incorporated the transit timing deviations into the light curve parameter fit in an iterative sense, similar to the procedure described by \cite{Lissauer11} for the transit times of the Kepler-11 planets: we first estimate the light curve parameters with an assumption of a linear ephemeris. We then fit for the epoch of each individual transit using these parameters in a manner that we describe below (with the exception of the epoch, which we allow to float for each transit). Finally, we shift the transits by their measured timing deviation, refold the light curve, and repeat the fit for the parameters. We repeat this process until all parameters converged, and found that it converged after only two iterations. We note that we employed only the subset of short cadence observations for fitting the light curve parameters themselves. The addition of the long cadence observations from Quarters 0-2 to the analysis would have contributed only a very modest improvement to our knowledge of the parameters.

We estimated the uncertainty in the parameters using the Markov Chain Monte Carlo (MCMC) method as follows, using a Metropolis-Hastings algorithm with Gibbs sampling \cite{Tegmark04}. We employ model light curves generated with the routines in \cite{Mandel02}, which depend upon the period $P$, the epoch $T_{c}$, the planet-to-star radius ratio $R_{p}/R_{\star}$, the ratio of the semi-major axis to the stellar radius $a/R_{\star}$, the inclination of the orbit $i$, and two quadratic limb-darkening coefficients $u_{1}$ and $u_{2}$. We assume an eccentricity of zero for the orbit, which we discuss in further detail in Section 3.2. We choose randomly one parameter, perturb it, and evaluate the $\chi^{2}$ of the solution. If the $\chi^{2}$ is lower, we accept the new parameter value. If the $\chi^{2}$ is higher, we evaluate the probability of accepting the jump as $p=e^{-\Delta\chi^{2}/2}$. We adjust the width of the distribution from which we randomly draw the jump sizes in each parameter until 20--25\% of jumps are executed in each of the parameters. We created five chains, each of length 10$^{6}$ points, where each of the chains is begun from a different set of starting parameters (each parameter is assigned a starting position that is +$3\sigma$ or -$3\sigma$ from the best-fit values). We discard the first 20\% of jumps from each chain to remove the transient dependence of the chain on the starting parameters.

After the first iteration of the MCMC procedure, we locate the best solution in the $\chi^{2}$ sense, and use these parameters to fit the individual times of transit for observations gathered in both short and long cadence observing mode. We fix all parameters with the exception of the center of transit time, which we allow to vary over a range of 30 minutes for each transit, centered on the predicted transit time from a linear ephemeris, in intervals of 6 seconds. In this case, we account for the finite integration time by taking a numerical average of an oversampled model (evaluated at 6 second intervals) over a period corresponding to the exposure time (58.8 seconds for short-cadence observations, and 29.5 minutes for long-cadence observations). We determine the center of transit time from the epoch associated with the minimum $\chi^{2}$ value, and the error from the range over which $\Delta\chi^{2}<1$ when compared to the minimum value. In practice, this results in asymmetric error bars for many of the individual transits. We also conduct an analysis to determine the contribution of quarter-by-quarter correlated noise to the transit time measurements, whereby we inject transits with the same light curve parameters as we derive for Kepler-19b into the \kepler\ light curve, and then fit for the transit times in an identical fashion to the authentic transits. Based upon a typical deviation of these times from the injected time (as compared to their error bars), we inflate the error bars for the short-cadence observations by factors of 1.32, 1.15, 1.45, 1.40, 1.18, and 1.20 for Quarters 3--8, respectively. These values are consistent with the larger scatter of the transit times in the latter quarters, as discussed in further detail in Section 6.1. The transits gathered in long cadence from Quarters 0--2 show errors that are consistent with Gaussian noise, and so we do not apply a scaling factor to these error bars. After two iterations of the steps described above, we found that the measured individuals transit times varied by less than 10 seconds between the two iterations. We fixed the average period to the one determined by the final fit to the transit times for the final MCMC analysis.

\begin{figure}
\begin{center}
 \includegraphics[width=8in,angle=270]{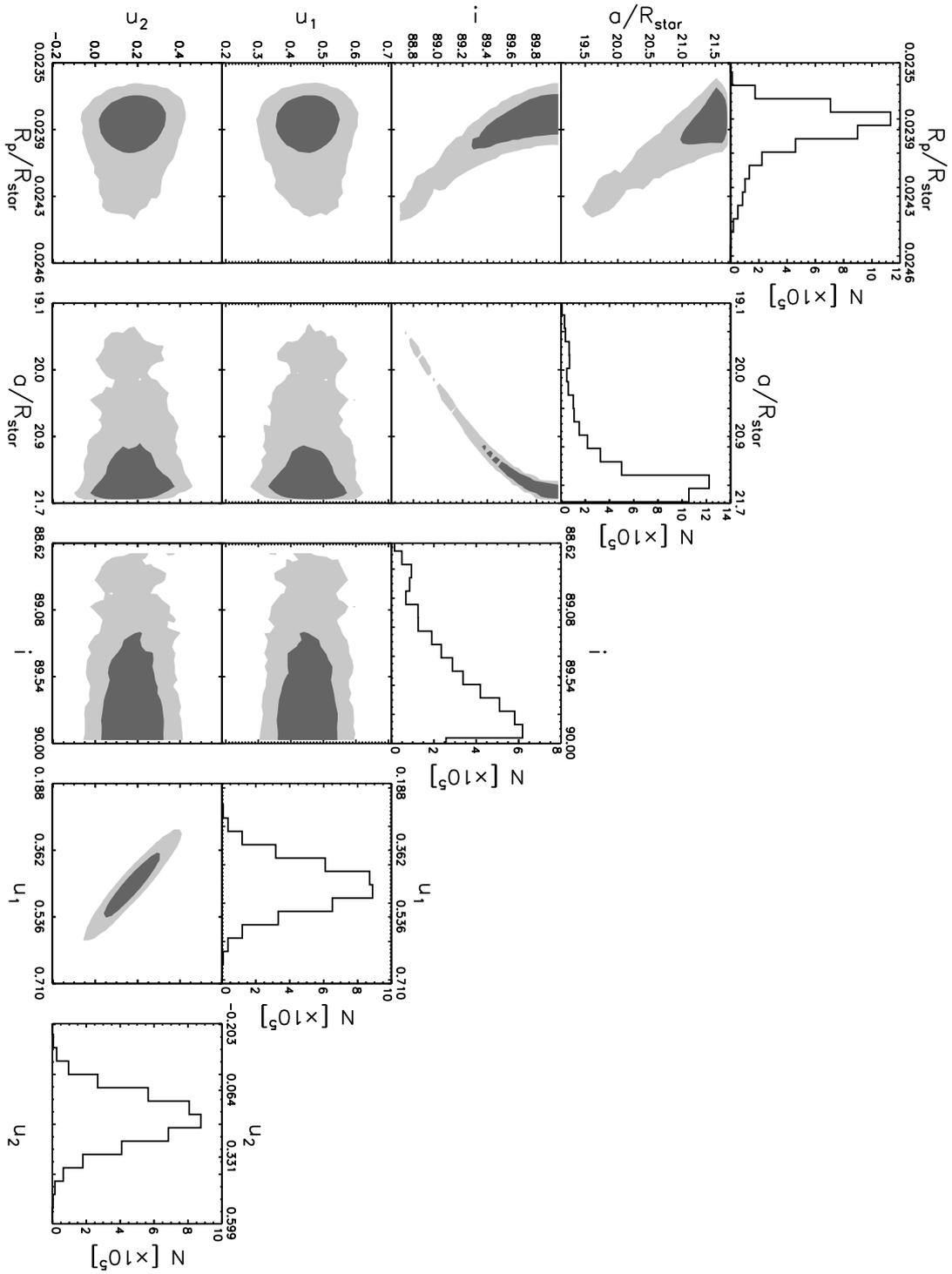} 
 \caption{Markov Chain Monte Carlo probability distributions for light curve parameters of Kepler-19b. The dark grey area encloses 68\% of the values in the chain, while the light grey area encloses 95\% of the values. We assign the range of values corresponding to 1$\sigma$ confidence from the area enclosing 68\% of the values nearest to the parameters associated with the minimum $\chi^{2}$ (as described in the text).}
  \label{fig:mcmc_results}
\end{center}
\end{figure}

\begin{figure}[h!]
\begin{center}
 \includegraphics[width=4in]{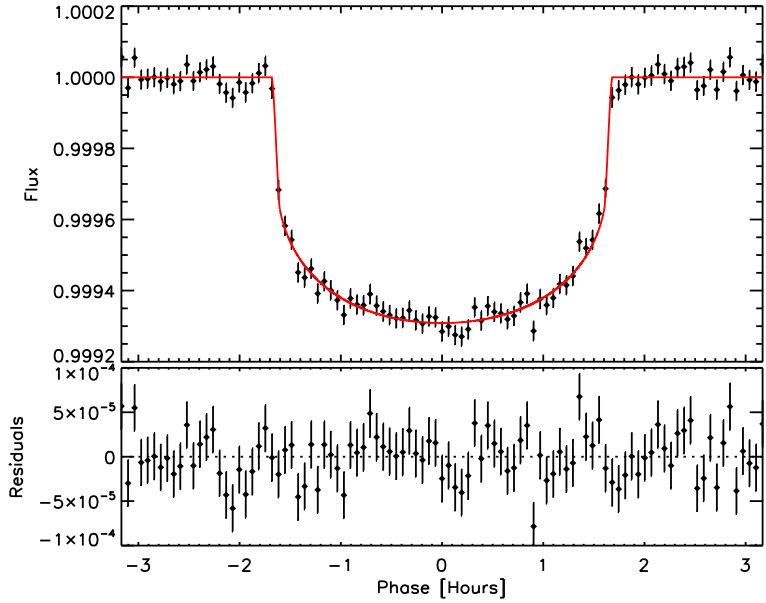} 
 \caption{{\slshape Kepler} transit light curve for the short-cadence observations of Kepler-19, centered on time of transit, with transit timing variations removed, and binned by a factor of 40 (binsize of 4 minutes). Overplotted in red is the best transit model light curve, with parameters given in Table \ref{tbl:params}. The bottom panels shows the residuals of the light curve, after the model is subtracted.}
  \label{fig:keplerfit}
\end{center}
\end{figure}

\begin{figure}[h!]
\begin{center}
 \includegraphics[width=4in]{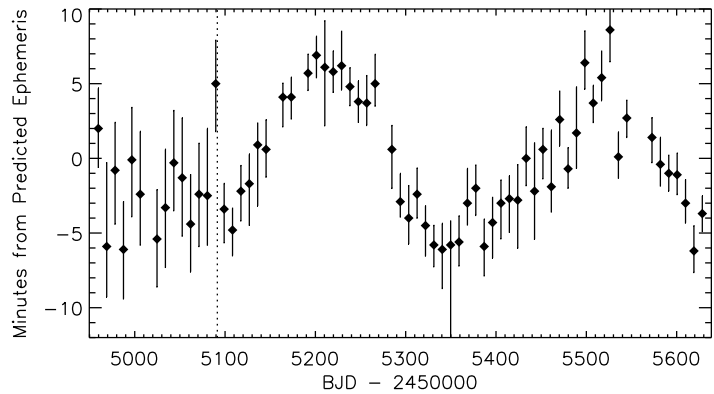} 
 \caption{{\slshape Kepler} transit times for Kepler-19b from Quarters 0-8, as compared to the best linear ephemeris model. The linear ephemeris we use to generate these O-C values is given in Table \ref{tbl:params}, and the individual transit times are given in Table \ref{tbl:times}. The demarcation between long cadence and short cadence observations is shown with a dotted line.}
  \label{fig:times}
\end{center}
\end{figure}

\begin{figure}[h!]
\begin{center}
 \includegraphics[width=4in]{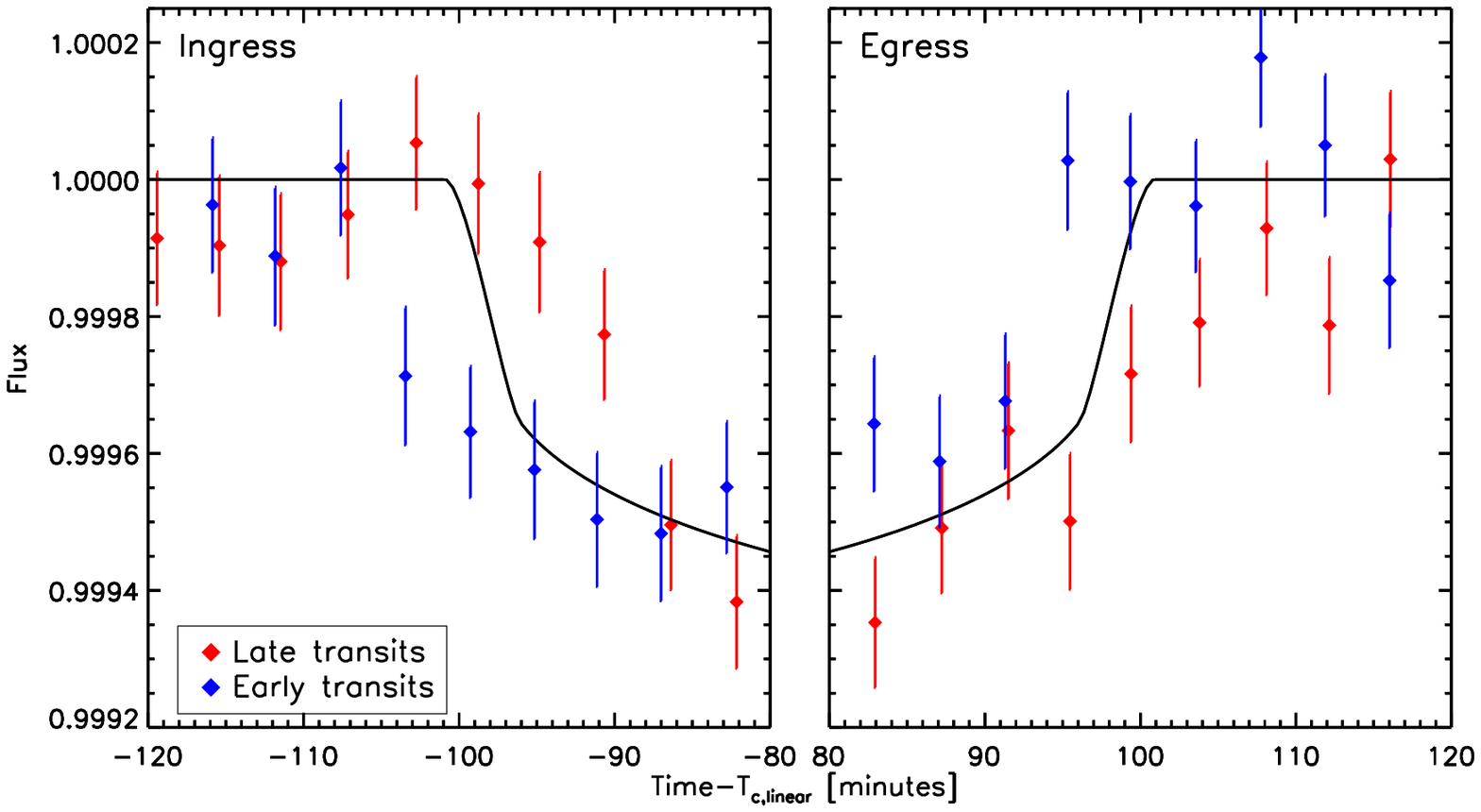} 
 \caption{A binned subset of five ``late'' transits of Kepler-19b (in red, comprised of individual transit numbers 26--30 listed in Table \ref{tbl:times}) and five ``early'' transits (in blue, comprised of transit numbers 39-43), with best linear transit model overplotted. At left, we show the binned transits centered on time of ingress, and at right, centered on time of egress.}
  \label{fig:ttvbyeye}
\end{center}
\end{figure}

In Figure \ref{fig:mcmc_results}, we show the MCMC correlations between all free parameters in the model fit, as well as the histograms corresponding to each parameter. In Figure \ref{fig:keplerfit}, we show the \kepler\ transit light curve for Kepler-19b (which is phased to the best period after shifting the transit times by the values given in Table \ref{tbl:times}), with the best-fit transit light curve overplotted. We report the best-fit parameters and uncertainties in Table \ref{tbl:params}. The range of acceptable solutions for each of the light curve parameters ($R_{p}/R_{\star}$, $a/R_{\star}$, and $i$) is determined as follows. We report the ``best'' solution from the set of parameters that minimize the $\chi^{2}$. The error bars are then given by the highest and lowest values that are within the 68\% of points closest in $\chi^{2}$ to the best value. We additionally calculate the transit duration, impact parameter, and ingress duration, using the formulae given in \cite{Seager03} and \cite{Winn10} to create the distribution in those quantities from the parameters in the MCMC chains. In some cases, the error bar is asymmetric (for $a/R_{\star}$ and $i$, which we expect from their asymmetrical MCMC distributions). We report the transit times, deviation from a linear ephemeris, and errors in Table \ref{tbl:times}. Figure \ref{fig:times} shows the individual timing deviations from the best linear ephemeris. In Figure \ref{fig:ttvbyeye}, we show a binned subset of ``late'' and ``early'' transits, comprised of five transits each (corresponding to numbers 26--30 and numbers 39--43 listed in Table \ref{tbl:times}), over which we plot a model generated with a linear ephemeris. The deviation of the transit times from the predicted $T_{c}$ of five minutes, which is equal to approximately one ingress time, is apparent.

We also performed the MCMC analysis with the two quadratic limb-darkening coefficients (LDCs), $u_{1}$ and $u_{2}$, fixed to theoretical values. We used the values for the effective stellar temperature $T_{\rm eff}$, metallicity [Fe/H], and surface gravity log(g) derived from our Spectroscopy Made Easy (SME) analysis of the Keck HIRES spectra (see Section 4.2): $T_{\rm eff}$=5541 $\pm$60 K, [Fe/H]=-0.13$\pm$0.06 dex, and log(g) of 4.59 $\pm$ 0.1 dex. In this case, the closest stellar model from the tables of theoretical limb-darkening coefficients generated for the \kepler\ bandpass by \cite{Prsa10} corresponded to a model with $T_{\rm eff}$=5500 K, log(g)=4.5, and [Fe/H]=0.0: these coefficients are $u_{1}$=0.5 and $u_{2}$=0.18. Our results for the planetary parameters were consistent with the values we obtained while allowing the limb-darkening coefficients to float, but the error bars were slightly smaller with fixed LDCs. In particular, the LDC derived from the light curve lie within 1$\sigma$ of the theoretical values for $u_{1}$ and $u_{2}$, and the largest deviation in the derived parameters is 1.4$\sigma$ for $R_{p}/R_{\star}$. The difference between the best $\chi^{2}$ values (between fixing the LDCs or allowing them to float) is approximately equal to five, which is roughly consistent with the addition of two degrees of freedom.


\subsection{Physical Parameters}
We based our procedure for constraining the mass, radius, and age of the host star on the method described by \cite{Torres08}. Using the metallicity determined from SME (described in Section 4.2), we created a set of stellar evolution models from the Yonsei-Yale (Y$^{2}$) isochrone series by \cite{Yi01}, with corrections from \cite{Demarque04}. We employed the interpolation software that accompanied that work, which accepts as inputs the age of the star, the iron abundance, and the abundance of $\alpha$-elements relative to solar (for which we assume the solar value), and outputs a grid of stellar isochrones corresponding to a range of masses. We evaluated a set of isochrones over an age range of  0.1 to 14 Gyr (at intervals of 0.1 Gyr) and in [Fe/H] in 60 equally spaced increments from $-3\sigma$ to $+3\sigma$ above and below the best-fit value of [Fe/H]=$-0.13\pm0.06$. We then performed a spline interpolation of each output table at a resolution of 0.005 $M_{\odot}$ in effective temperature $T_{\rm eff}$, the log of the surface gravity log(g), and the stellar luminosity $L_{\star}$. We evaluate the physical radius corresponding to each stellar model via log(g) and the mass of the star ($g=GM_{\star}/R_{\star}^{2}$), though it is also possible to convert to physical radius using the model stellar luminosity and effective temperature (assuming $L_{\star}=4\pi R_{\star}\sigma T^{4}$); in practice these conversions give identical results.

Rearranging Kepler's version of Newton's third law in the manner employed by \cite{Seager03}, \cite{Sozzetti07} and \cite{Torres08}, we convert the period (derived from photometry), and the radius and mass of the host star (from isochrones) to a ratio of the semi-major axis to the radius of the host star, $a/R_{\star}$:
 
\begin{equation}
\frac{a}{R_{\star}}=\left(\frac{G}{4\pi^{2}}\right)^{1/3}\frac{P^{2/3}}{R_{\star}}(M_{\star}+M_{p})^{1/3},
 \label{eq:kepler}
\end{equation}
\noindent where we assume that $M_{p}$ is negligible when compared to the mass of the host star, and that the orbit is circular. Using the MCMC sequence of $a/R_{\star}$  and generating a series of Gaussian random realizations of [Fe/H] and $T_{\rm eff}$ using the values and error bars derived from spectroscopy, we locate the best isochrone fit at each realization using the $\chi^{2}$ goodness-of-fit 
\begin{equation}
\chi^{2}=\left(\frac{\Delta a/R_{\star}}{\sigma_{a/R_{\star}}}\right)^{2}+\left(\frac{\Delta T_{\rm eff}}{\sigma_{T_{\rm eff}}}\right)^{2}+\left(\frac{\Delta[\mbox{Fe/H}]}{\sigma_{\mbox{[Fe/H]}}}\right)^{2}.
\label{eq:chisquared}
\end{equation}
\noindent Using the output of the MCMC chain of $a/R_{\star}$ ensures that correlations between parameters, which are preserved in the chain, are properly incorporated into our estimate of the stellar parameters.

We then assign a weight to the likelihood of each stellar model in the chain by applying a prior for the initial mass function (IMF) that assumes a Salpeter index \citep{Salpeter55}. The number of stars of each mass and age, per 1000 stars, is generated by the interpolation software provided by \cite{Yi01} for several IMF assumptions, including the Salpeter IMF. We designate the weight assigned to each stellar model in the chain by normalizing to the highest IMF value within the sample: in practice, the weights vary from 0.2 to 1 (from the least to most likely). We then incorporate this likelihood by discarding members of the chain according to their weight, where the weight is equal to the likelihood of remaining in the chain. About 50\% of the original chain remains intact after this stage. The value for each stellar parameter is then assigned from the median of this weighted distribution, with the formal error bars assigned from the nearest 68\% of values above and below the median. We find $M_{\star}$= 0.936 $\pm$ 0.040 $M_{\odot}$, $R_{\star}$= 0.850 $\pm$ 0.018 $R_{\odot}$, and an age = 1.9 $\pm$ 1.7 Gyr. These uncertainties exclude possible systematic uncertainties in the stellar models. Using the modified MCMC chain in both stellar radius and $R_{p}/R_{\star}$ to determine the physical radius of the planet, we find $R_{p}$=2.209 $\pm$ 0.048 $R_{\oplus}$. The IMF prior changes the final answer by less than 1$\sigma$ for all physical parameters, and by 0.1$\sigma$ in the case of the planetary radius.  

We note that we also recorded the log(g) of the best stellar model for each realization of $T_{\rm eff}$, [Fe/H], and $a/R_{\star}$, as described above. From this analysis, we find a log(g) of 4.54$\pm$0.02, which is consistent with the value inferred from spectroscopy of 4.59$\pm$0.10. We conclude that the assumption of zero orbital eccentricity for the transiting planet is consistent with the value of log(g) measured spectroscopically. However, using the analytic formulae presented in \cite{Carter08}, the derived log(g) would vary by only 0.04 dex if the eccentricity were as high as 0.15 (which \citealt{Moorhead11} found was typical for the sample of \kepler\ Objects of Interest), which is well below our measured uncertainty on log(g).

\begin{deluxetable}{rr}
\tablenum{1}
\tablecaption{Star and Planet Parameters for Kepler-19}
\label{tbl:params}
\tablewidth{0pt}
\tablehead{
\colhead{Parameter}    & \colhead{Value}\\}
\startdata
Kepler-19 [star]&\\
\hline
Right ascension & 19h21m40.99s\\
Declination & +37d51m06.5s\\ 
$\mbox{[Fe/H]}$ & -0.13$\pm$0.06 \\
log(g) [cgs] & 4.59$\pm$0.10 \\
$T_{\rm eff}$ [K] & 5541$\pm$60 \\
v sin $i$ [km s$^{-1}$] & $<2$ \\
 $M_{\star}$ [$M_{\odot}$]  &  0.936$\pm$0.040\\
$R_{\star}$ [$R_{\odot}$]  &  0.850$\pm$0.018\\
Age [Gyr] & 1.9$\pm$1.7 \\
\hline
Kepler-19b &\\
\hline
Period [days]   &  9.2869944$\pm$0.0000088\tablenotemark{a}\\
$T_{c}$ [BJD-2450000]   &    4959.70597$\pm$0.00036 \\
$R_{p}/R_{\star}$   &      0.02379$\pm$0.00012\\
$a/R_{\star}$    &     21.59$^{+0.15}_{-0.37}$\\
$i$ [degrees]  &       89.94$^{+0.06}_{-0.44}$\\ 
$u_{1}$ & 0.466$\pm$0.061\\
$u_{2}$ &    0.155$\pm$0.097\\
Impact parameter, $b$ & 0.02$^{+0.16}_{-0.02}$  \\
Transit duration, $\tau$ [min]  &  201.91$\pm$0.47 \\
Ingress duration, $\tau_{ing}$ [min] & 4.70$^{+0.18}_{-0.57}$\\ 
$R_{p}$ [$R_{\oplus}$] & 2.209$\pm$0.048 \\
$M_{p}$ [$M_{\oplus}$] & $<$20.3\tablenotemark{b}\\
\hline
Kepler-19c & \\
\hline
Period [days] & $<$160\tablenotemark{c} \\
$M_{p}$ [$M_{\rm Jupiter}$] & $<$6.0\tablenotemark{c}\\
\enddata
\tablenotetext{a}{The period and $T_{c}$ values for Kepler-19b are determined from a linear fit to the transit times.}
\tablenotetext{b}{The upper limit on the mass of Kepler-19b is determined from the radial velocity analysis in Section 4.2.}
\tablenotetext{c}{The upper limit on the mass and period of Kepler-19c is described in the text of Section 6.1.2.}
\end{deluxetable}

\section{Follow-up Observations}

\subsection{Reconnaissance Spectroscopy}
\label{sec:recon}
We gathered reconnaissance spectra of Kepler-19 on UT 2009 August 05 (orbital phase 0.596), 2009 August 29 (phase 0.161) and 2010 October 1 (phase 0.017) with the Tull coude spectrograph of the McDonald Observatory 2.7m telescope.  We used these high resolution (R=60,000) spectra to verify the Kepler Input Catalog stellar classification and to search for evidence of any secondary stellar spectrum or binary orbital motion.  We cross-correlated the spectra against a library of synthetic stellar spectra as described by \cite{Batalha11}.  The spectra did not show any evidence of a secondary spectrum, and the absolute radial velocities, which cover both orbital quadratures, agree at the 0.75 km/s level.  These spectra gave the best match to the synthetic templates with $T_{\rm eff}$ = 5750 K, log(g)= 4.5, and v sin $i$ = 2 km s$^{-1}$, for an assumed solar metallicity.  The height of the cross-correlation peaks was 0.98 for all of the spectra, indicating an excellent fit to the stellar template spectra.

\subsection{High-resolution Spectroscopy}

Between 2009 October 29 and 2011 June 10 we gathered 8 high-resolution spectra of Kepler-19 with the Keck HIRES spectrometer \citep{Vogt94}. With these spectra, we conducted an analysis to determine the stellar parameters. We compared a high-resolution template spectrum to stellar models, generated with the spectral synthesis package Spectroscopy Made Easy (SME; \citealt{Valenti96, Valenti05}). We determine the effective temperature $T_{\rm eff}$ of the host star of 5541 $\pm$ 60 K, a metallicity [Fe/H] of $-$0.13 $\pm$ 0.06 dex, a log(g) of 4.59 $\pm$ 0.1 dex, and a v sin $i$ $<$ 2 km s$^{-1}$. We comment briefly here on the stellar activity. We find a value of the ratio of emission from the Ca II H and K lines to the total bolometric emission of log($R_{HK}$)=-4.95$\pm$0.05. The $R_{HK}$ value is derived from a Mt. Wilson style S-value of 0.174$\pm$0.007 \citep{Isaacson10}. The log($R_{HK}$) value is low for main sequence stars of this temperature and is consistent with the slow stellar rotation we infer from the measured v sin $i$ of $<$2 km s$^{-1}$. If we assume rigid body rotation of the star and a stellar spin axis aligned with the orbital spin axis of the planet, we find a lower limit on the stellar rotation period of 22 days. The rotation period derived from the $R_{HK}$ value is 32 days \citep{Noyes84} which along with the lack of emission in the core of the Ca II H and K lines leads us to conclude that the star is relatively inactive.

We further used the spectra to derive estimates of the stellar radial velocity. The spectra were gathered with the same configuration of HIRES, described in \cite{Marcy08}, which was demonstrated to yield typical precisions of 1.0 to 1.5 m s$^{-1}$ on nearby FGK stars. This method relies on the use of an iodine cell placed in front of the beam, which superimposes the iodine spectrum on the stellar spectrum with an identical instrumental profile. For each 100 pixel section of the spectrum, the iodine and stellar spectral lines are fit simultaneously. For the set of these observations, this treatment yields a typical internal error estimate on individual radial velocities of 1.5 m s$^{-1}$. We note also the use of the ``C2 decker'' entrance aperture for all of these observations, which technique is described in greater detail in \cite{Batalha11}, and enables sky subtraction (as compared to the ``B5 decker'', for which sky subtraction is not possible). We list the measured radial velocities, with associated error bars (excluding stellar jitter), in Table \ref{tbl:rv}. We gathered twelve additional observations prior to the ones listed in Table \ref{tbl:rv} but these were observed with the B5 configuration, and had a much higher scatter (15 m s$^{-1}$ as opposed to 4 m s$^{-1}$). For this reason, we excluded them from the analysis. 

\begin{figure}[h!]
\begin{center}
 \includegraphics[width=5in]{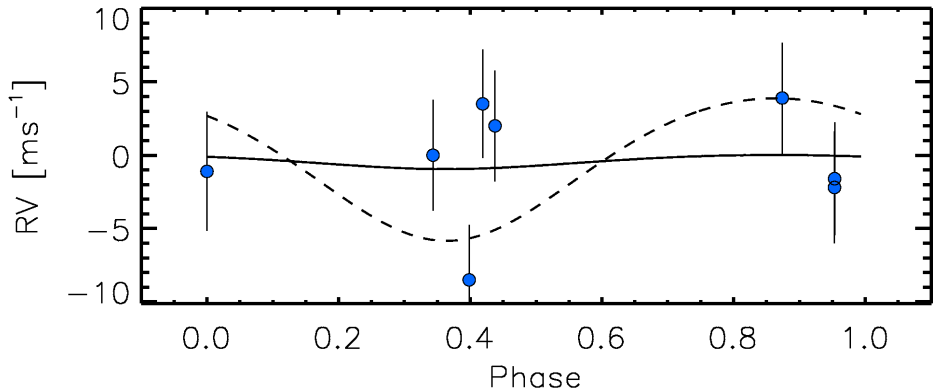} 
 \caption{Measured radial velocities of Kepler-19, as a function of phase (assuming the orbital period and epoch of Kepler-19b, as stated in Table \ref{tbl:params}). We have depicted two radial velocity models with zero eccentricity, the first (dashed) corresponding to a planetary mass at the $3\sigma$ upper limit of  15.2 $M_{\oplus}$, and the second (solid) corresponding to the most likely amplitude of 0.5 m s$^{-1}$ (or 1.6 $M_{\oplus}$). The error bar depicted include the effects of stellar jitter, which we conclude are near 4 m s$^{-1}$.}
  \label{fig:rv}
\end{center}
\end{figure}

We determine the upper mass limit on Kepler-19b from the radial velocities as follows. We employ the Bayesian MCMC technique described in \cite{Gregory07} to fit a radial velocity model to the observations. The free parameters in the model are the semi-amplitude velocity $K$ of the star, the zero-point velocity $\gamma$, the eccentricity $e$ of the planetary orbit, the argument of perihelion $\omega$, and a stellar jitter term. The orbital period and transit epoch are also free parameters, however the precision of the Gaussian priors we place on them from the light curve analysis (see Table \ref{tbl:params}) effectively fixes their values. Additionally, we fix the inclination of the orbit to the value measured from the light curve of 89.94$^{\circ}$. We first fit a model, assuming a circular orbit. In this case, we infer a stellar jitter contribution of 4.1 $\pm$ 1.7 m s$^{-1}$, and 1, 2 and 3$\sigma$ upper limits on the semi-amplitude of 1.4, 3.3, and 4.9 m s$^{-1}$, respectively (these values are derived from integrating over the posterior distribution of semi-amplitude until 68\%, 95\%, and 99.73\% of the area is enclosed). Applying the semi-amplitude toward a mass upper limit, we determine a 3$\sigma$ upper limit on the mass of $<15.2$ $M_{\oplus}$, if the orbit is circular (these upper limits are 4.3 and 10.2 $M_{\oplus}$ at 1 and 2$\sigma$ confidence). The most likely fit (depicted as the solid line in Figure \ref{fig:rv}) has an amplitude of 0.5 m s$^{-1}$ (corresponding to a mass of 1.6 $M_{\oplus}$), but this value is well below our detectability threshold. For comparison, we also show representative circular orbit at the 3$\sigma$ upper limit for semi-amplitude (dashed line in Figure \ref{fig:rv}).

We also address the possibility of a non-zero eccentricity. If the orbit were significantly eccentric, then the transit duration of Kepler-19b would deviate from the predicted duration for the edge-on circular orbit scenario, unless the argument of perihelion conspired to mimic the circular transit duration. As eccentricity increases, there is an increasingly narrow range of $\omega$ that matches the transit duration of a circular orbit. As we demonstrate in Section 3, the assumption of a circular orbit is consistent with the well-constrained low impact parameter (high inclination) measured from the the \kepler\ light curve, as illustrated by the short ingress and egress times. In addition the log(g) inferred from the light curve analysis assuming zero eccentricity agrees with the independent spectroscopically determined value. We note that the planet is too small to constrain the eccentricity from secondary eclipse observations; we discuss this possibility further in Section 6.3.  Therefore, we elected to impose a prior on $e$ and $\omega$ from our knowledge of the transit duration, as follows. The ratio $\tau$ of the transit duration for an eccentric orbit to the transit duration for a circular orbit can be approximated by the following expression of the eccentricity and argument of perihelion \citep{Burke08}:

\begin{equation}
\tau=(1-e^{2})^{1/2}/(1+e\cos(\omega-\pi/2)).
 \label{eq:tau}
\end{equation}
\noindent For each element of the MCMC chain, for which we now vary $e$ and $\omega$, we evaluate the transit duration ratio $\tau$. We then assign a flat prior on $\tau$, which is equal to one for 0.7$<\tau<$1.3 (that is, the transit duration is within 30\% of circular) and zero otherwise. Applying this transit duration prior during the radial velocity parameter estimates, we find a negligibly smaller estimate for the stellar jitter than the circular orbit case and semi-amplitude upper limits (again, with 1, 2, and 3$\sigma$ confidence respectively) of 2.0, 9.2, and 23.5 m s$^{-1}$. These semi-amplitudes are associated with mass upper limits of 5.5, 20.3, and 50.3 $M_{\oplus}$, with the same stated confidences. Given the radius value for the planet determined from the \kepler\ photometry and the $2\sigma$ upper mass limit of 20.3 $M_{\oplus}$, we find an upper limit on the density of Kepler-19b of 10.4 g cm$^{-3}$. We comment further on the possible composition of the planet, given these upper limits, in Section 6.4. 
\begin{deluxetable}{rrr}
\tablenum{3}
\tablecaption{Relative Radial Velocities for Kepler-19}
\label{tbl:rv}
\tablewidth{0pt}
\tablehead{
\colhead{HJD}         & \colhead{RV}     & \colhead{Unc.}  \\
\colhead{-2450000}   & \colhead{m s$^{-1}$}  & \colhead{m s$^{-1}$}  }
\startdata
5134.805   &    0.9   &   2.1 \\
5320.113   &    0.4   &   1.7 \\
5402.956   &    5.9   &   1.5 \\
5407.827   &   -6.5   &   1.4 \\
5412.981   &   -0.2  &    1.6 \\
5435.879   &    5.5   &   1.3 \\
5723.074   &    2.0   &   1.5 \\
5723.950   &    4.0   &   1.5 \\
\enddata
\tablecomments{Uncertainties do not include stellar jitter, which is likely to be near 4 m s$^{-1}$.}
\end{deluxetable}

\subsection{Adaptive Optics Imaging}
\label{sec:ao}

\begin{figure}[h!]
\begin{center}
 \includegraphics[height=5in,angle=90]{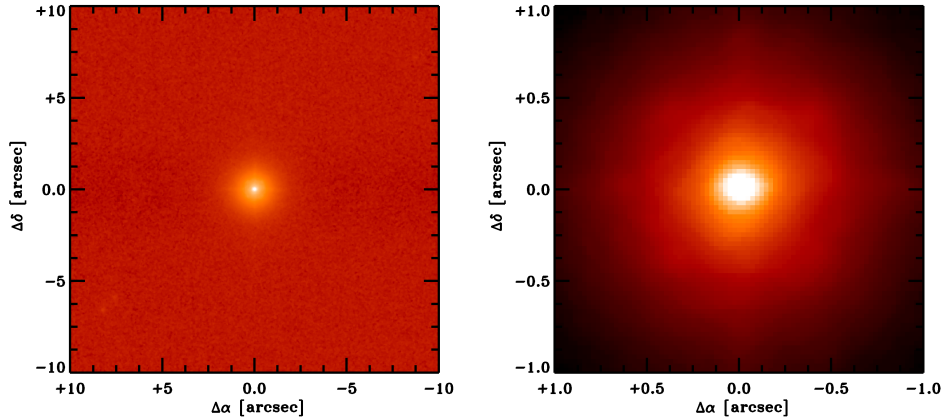} 
 \caption{$J$-band adaptive optics image of the neighborhood of Kepler-19, within 10 arcseconds (left) and 1 arcsecond (right).}
  \label{fig:ao1}
\end{center}
\end{figure}

We gathered adaptive optics images in $J$ band of Kepler-19 on UT 24 September 2009, using the PHARO near-infrared camera \citep{Hayward01} on the the Hale 200 inch telescope on Mt. Palomar, CA. \cite{Troy00} give a complete description of the Palomar adaptive optics system. We employed a dither pattern for these observations similar to the technique described in \cite{Batalha11}, although we used an exposure time of 4.25 seconds. In Figure \ref{fig:ao1}, we show the local neighborhood of Kepler-19 within 10 arcseconds. 

\begin{figure}[h!]
\begin{center}
 \includegraphics[height=3in]{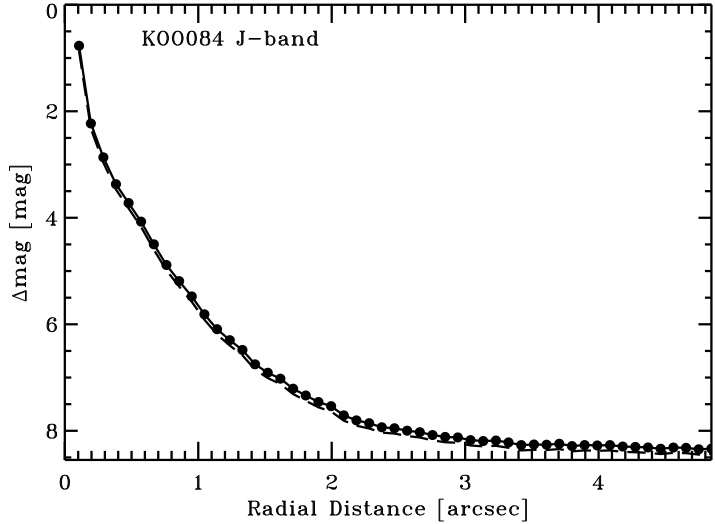} 
 \caption{The sensitivity limits to additional point sources in the neighborhood of Kepler-19b as a function of radial distance from the primary target.  The filled circles represent the J-band limits and each point represents a step in FWHM away from the primary target centroid peak.  The dashed line underneath represents the $J$-band limits converted to \kepler\ magnitude limits if a star were to have a nominal $Kp$-$J$ color, as described in the text.}
  \label{fig:ao2}
\end{center}
\end{figure}

We assess our sensitivity to additional sources using a similar procedure to that described by \cite{Batalha11}. We inject fake sources near the target star at random position angles, using steps in magnitude of 0.5 mag and varying the distance from the target star in increments of 1.0 FWHM of the point-spread function (PSF). We then attempt to identify the injected sources with the DAOPhot routine \citep{Stetson87} and also by eye, and set our sensitivity limit, as a function of distance, at the magnitude where we are able to recover the injected sources. The limit in $\Delta m$ as a function of distance from the target star is shown in Figure \ref{fig:ao2}. We then convert the $\Delta m$ sensitivity limit in $J$ band to a limit in \kepler\ magnitudes, by assuming a nominal \kepler\ magnitude-$J$ color (using the value derived from a magnitude-limited sample of $Kp$-$J$=1.28$\pm$0.52 mag). We do not detect any additional sources within our sensitivity limits in the neighborhood of Kepler-19.

\subsection{Speckle Imaging}
\begin{figure}[h!]
\begin{center}
 \includegraphics[height=6in]{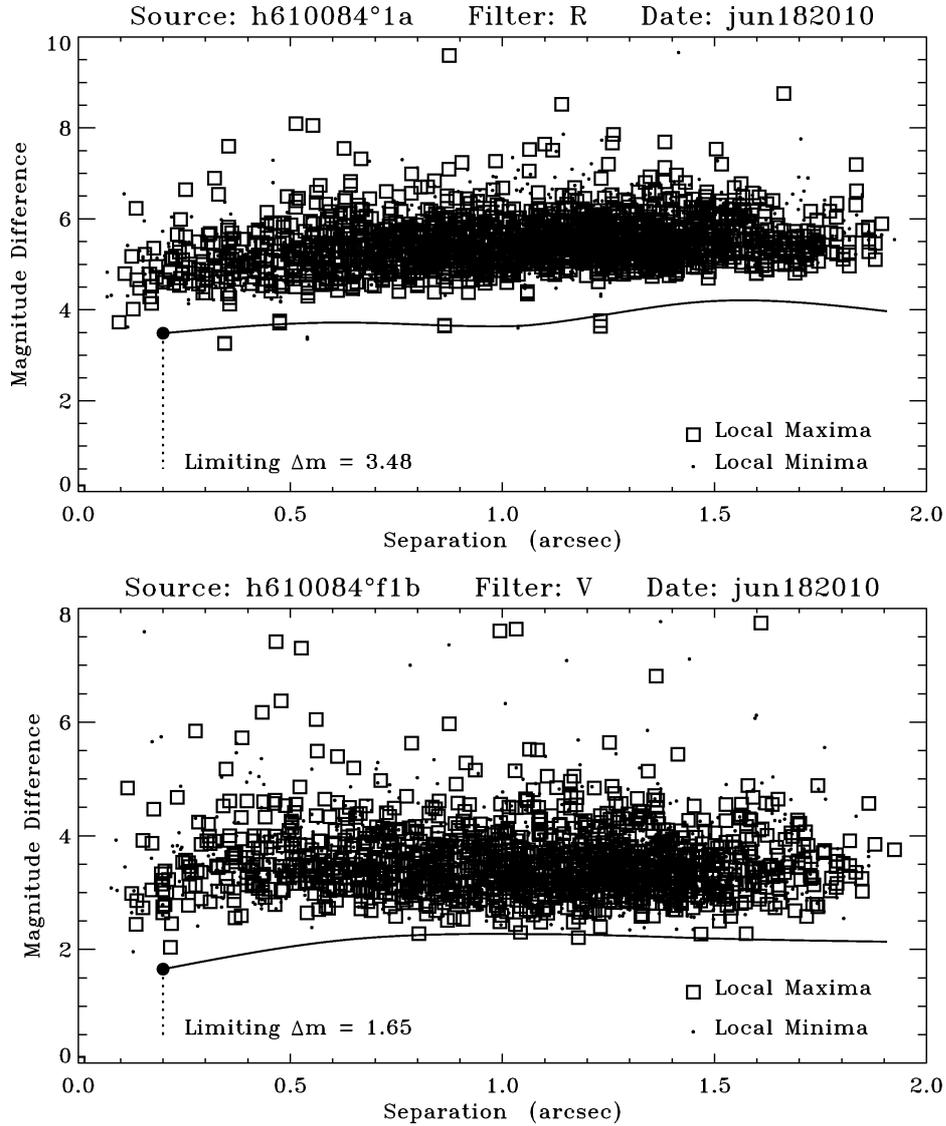} 
 \caption{{\it Top panel:} $R$ band speckle sensitivity curve of Kepler-19. The magnitude difference between the target star and local extrema in the background are denoted by squares (local maxima) and points (local minima). The solid line denotes a flux that is 5$\sigma$ brighter than the mean background brightness (where $\sigma$ is the standard deviation of the extrema in the background), where we could confidently detected an additional source. No companions are detected within 1.8 arcseconds of the target star to a depth of 4 magnitudes. {\it Bottom panel:} $V$ band speckle sensitivity curve of Kepler-19. No companions are detected within 1.8 arcseconds of the target star to a depth of 2 magnitudes.}
  \label{fig:speckle}
\end{center}
\end{figure}

We gathered speckle images of Kepler-19 using filters in both $R$ and $V$ band on UT 18 June 2010, using the Differential Speckle Survey Instrument located at the WIYN telescope (DSSI, \citealt{Horch09}) A detailed discussion of the recent upgrades to DSSI is presented in \cite{Horch10}, and summary of the speckle imaging survey of \kepler\ candidates, and those reduction procedures, is given by \cite{Howell11}. We assess our sensitivity to the presence of additional stars near the \kepler\ target star as a function of angular distance. For concentric rings of varying radius, centered on the target star, we determine the magnitude difference between the target star itself and the local extrema of the sky background. Figure \ref{fig:speckle} shows the results of this procedure in both $R$ and $V$ band. We find that we would have detected a companion at a distance of 0.2'' with a difference in magnitude smaller than $\Delta m$=3.48 in $R$ band, and a companion at a distance of 0.25'' with $\Delta m<$1.65 in $V$ band. Here again, as in the adaptive optics images, we detect no additional sources nearby to the \kepler\ target star.

\section{Planetary Validation of Kepler-19b}

\subsection{Photocenter Tests}
\label{sec:centroid}

We use two methods to search for false positives due to background eclipsing binaries, based on examination of the pixels in the aperture of Kepler-19: direct measurement of the source location via difference images, and inference of the source location from photocenter motion associated with the transits.  We employ two methods because of their different vulnerabilities to systematic bias; when the methods agree, we have increased confidence in their result.

Difference image analysis \citep{Torres11} takes the difference between average in-transit pixel images and average out-of-transit images.  A fit of the Kepler pixel response function (PRF; \citealt{Bryson10})  to both the difference and out-of-transit images directly provides the location of the transit signal relative to the host star.  We measure difference images separately in each quarter, and estimate the transit source location as the robust uncertainty-weighted average of the quarterly results.

We measure photocenter motion by computing the flux-weighted centroid of the pixels in the optimal aperture, plus a 
one-pixel halo in every cadence, generating a centroid time series for row and column. We fit the modeled transit to the whitened centroid time series transformed into sky coordinates.  We perform a single fit for all quarters, and then infer the source location by scaling the difference of these two centroids by the inverse of the flux as described in \cite{Jenkins10b}.  

\begin{figure}[htb]
\centering
\includegraphics[width=6in]{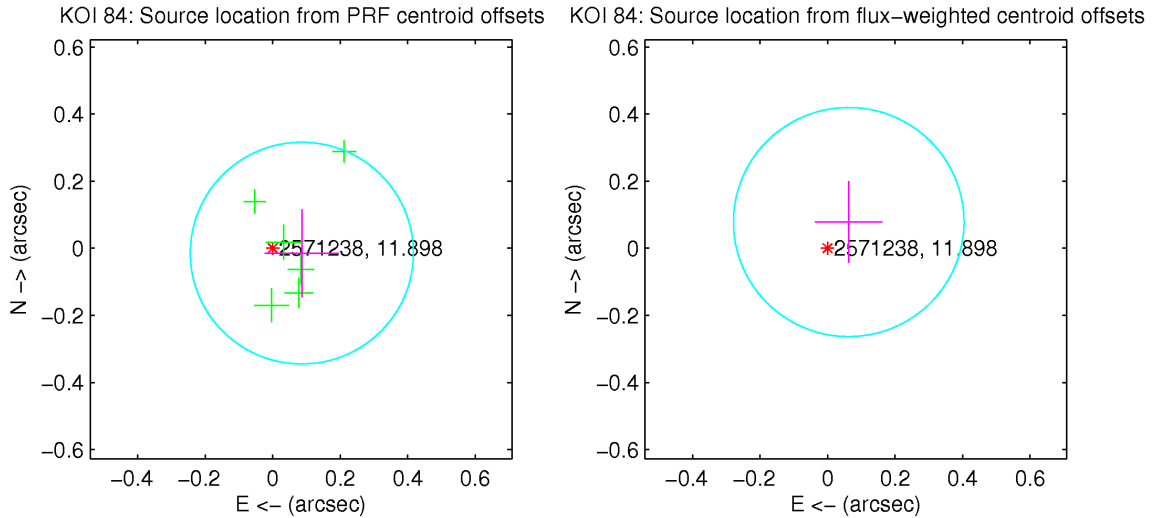} \\
\caption{Quarterly and average reconstructed transit source locations relative to Kepler-19.  {\it Left:} The green crosses show the individual quarter measurements using the difference image technique, and the magenta cross shows the uncertainty-weighted robust average of the quarterly results.  {\it Right:} The magenta cross shows the transit source location reconstructed from the multi-quarter fit of the transit signal to the centroid motion. The length of the crosses show the  $1 \sigma$ uncertainty of each measurement in RA and Dec.  The circles show the $3 \sigma$ circle around the 
average source location.  The location of Kepler-19 is shown by the red asterisk along with its Kepler ID and Kepler magnitude. }
\label{fig:centroid}
\end{figure}

The source as determined by the difference image method is offset from the nominal location of Kepler-19, as given in the \kepler\ Input Catalog, by 0.09 $\pm$ 0.11 arcsec $= 0.80 \sigma$. The source as determined by the flux-weighted centroid method is offset from Kepler-19 by 0.10 $\pm$ 0.11 arcsec $= 0.88 \sigma$.  The location of the offsets is shown for both methods in Figure~\ref{fig:centroid}.  Both methods show that the observed centroid location is consistent with the transit occurring at the location of Kepler-19.

\subsection{\it{Spitzer} Observations}
Warm \spitzer\ observations in the near-infrared can also prove useful toward validating \kepler\ candidates, as shown for Kepler-10c \citep{Fressin11}. Unless a putative blend scenario is comprised of stars of nearly identical color, the transit depth in a blend scenario will depend upon the wavelength at which it is observed. Conversely, an authentic transiting planet will produce an achromatic transit depth.

We gathered observations using the Infrared Array Camera (IRAC) \citep{Fazio04} on Warm {\it Spitzer} at 4.5~$\mu$m of two consecutive transits of Kepler-19: one on UT 29 June 2010, and one on UT 9 July 2010. Both observations span 8 hours, centered on the 3.5-hour-long transit. We gathered the observations using the full-array mode of IRAC, with an integration time of 26.8 s/image. We employed the techniques described in \cite{Agol10} for the treatment of the images before photometry. We first converted the Basic Calibrated Data products from the {\it Spitzer} IRAC pipeline (which applies corrections for dark current, flat field variations, and detector non-linearity) from mega-Janskys per steradian to data number per second, using 0.1469 MJy$\cdot$sr$^{-1}$ per DN s$^{-1}$, and then to electrons per second, using the gain of 3.71 $e$ DN$^{-1}$. We identified cosmic rays by performing a pixel-by-pixel median filter, using a window of 10 frames. We replace pixels that are $>4\sigma$ outliers within this window with the running median value. We also corrected for a striping artifact in some of the Warm {\it Spitzer} images, which occurred consistently in the same set of columns, by taking the median of the pixel values in the affected columns (using only rows without an overlying star) and normalizing this value to the median value of neighboring columns. 
 
We estimate the position of the star on the array using two techniques. First, we employed a flux-weighted sum of the position within a circular aperture of 3 pixels (we tested whether the size of this aperture made a difference by increasing the size to 4 pixels and repeating the analysis: we found that the position estimates were nearly identical). Additionally, we fit a Gaussian to the core of the PSF using the IDL routine \verb=GCNTRD= (again using apertures of 3 and 4 pixels, and finding no significant difference between them). We then performed aperture photometry on the images, using both estimates for the position and variable aperture sizes between 2.1 and 4.0 pixels, in increments of 0.1 pixels up to 2.7 pixels, and then at 3.0 and 4.0 pixels. We decided to use the position estimates using a flux-weighted sum at an aperture of 2.6 pixels, which minimized the out-of-transit RMS. Alternatively, using the positions derived with \verb=GCNTRD=, using a slightly smaller aperture, or using a slightly larger aperture, changed the RMS by only a few percent at the most.

We remove the effect of the IRAC intrapixel sensitivity variations, or the ``pixel-phase'' effect (see eg. \citealt{Charbonneau05, Knutson05}) using two techniques. With the first technique, we assume a polynomial functional form for the intrapixel sensitivity (which depends upon the $x$ and $y$ position of the star on the array). We denote the transit light curve $f$ (which depends upon time), and we hold all parameters constant except for the transit depth. We use the light curve software of \cite{Mandel02} to generate the transit models. The model for the measured brightness $f'(x,y)$ is given by:

\begin{equation}
f'=f(t,R_{p}/R_{\star})\cdot(b_{1}+b_{2}(x-\bar{x})+b_{3}(x-\bar{x})^{2}+b_{4}(y-\bar{y})+b_{5}(y-\bar{y})^{2}),
\label{eq:spitzerpoly}
\end{equation}

where we include all of the observations (both in- and out-of-transit) to fit the polynomial coefficients and the transit depth simultaneously.

We included cross-terms in $x$ and $y$, as well as higher order terms, but found that did not substantially decrease the RMS error of the out-of-transit residuals after the sensitivity function is divided from the flux. We fit for the polynomial coefficients $b_{1}$ through $b_{5}$ using a Levenberg-Marquardt $\chi^{2}$ minimization. We also performed an MCMC analysis to fit the polynomial coefficients to determine whether fitting for the transit depth was degenerate with any other free parameters, and determined that about 20\% of the error in the best-fit transit depth is due to a degeneracy with the strongest polynomial dependence of the intrapixel sensitivity, which is the linear coefficient in $y$. However, the Spitzer light curve contains significant correlated noise even after the best intrapixel sensitivity model is removed. We incorporate the effect of remaining correlated noise with a residual permutation analysis of the errors as described by \cite{Winn08}, wherein we find the best-fit model $f'$ to the light curve as given by Equation \ref{eq:spitzerpoly}, subtract this model from the light curve, shift the residuals by one step in time, add the same model back to the residuals, and refit the depth and pixel sensitivity coefficients. We wrap residuals from the end of the light curve to the beginning, and in this way we cycle through every residual permutation of the data. We determine the best value from the median of this distribution, and estimate the error from the closest 68\% of values to the median. We gathered 4.5 hours of observations outside transit, which is sufficient to sample the systematics on the same timescale as the 3.5 hour transit. Using the residual permutation method on the light curve treated with a polynomial, we find a best-fit transit $R_{p}/R_{\star}$ to be $0.0226\pm0.0039$ for the visit on 29 June, which is consistent with the best solution using MCMC, although the error bars are inflated by 40\% when compared to the MCMC error bars. For the visit on 9 July, we find $0.0280\pm 0.0027$ with the rosary-beading analysis; these error bars are 20\% larger than the corresponding MCMC error bars. The larger error on the transit depth measured on the first visit is attributable to the larger area on the pixel over which the star wanders during the observations: the smaller this area, the better we are able to fit the polynomial sensitivity model. While the extent of the pointing oscillations in the $x$ direction are comparable between the two visits (0.1 pixels), they differ substantially in the $y$ direction. The star wanders 0.15 pixels in $y$ on the 29 June visit and 0.08 pixels in $y$ on the 9 July visit. 

\begin{figure}[h!]
\begin{center}
 \includegraphics[width=6in]{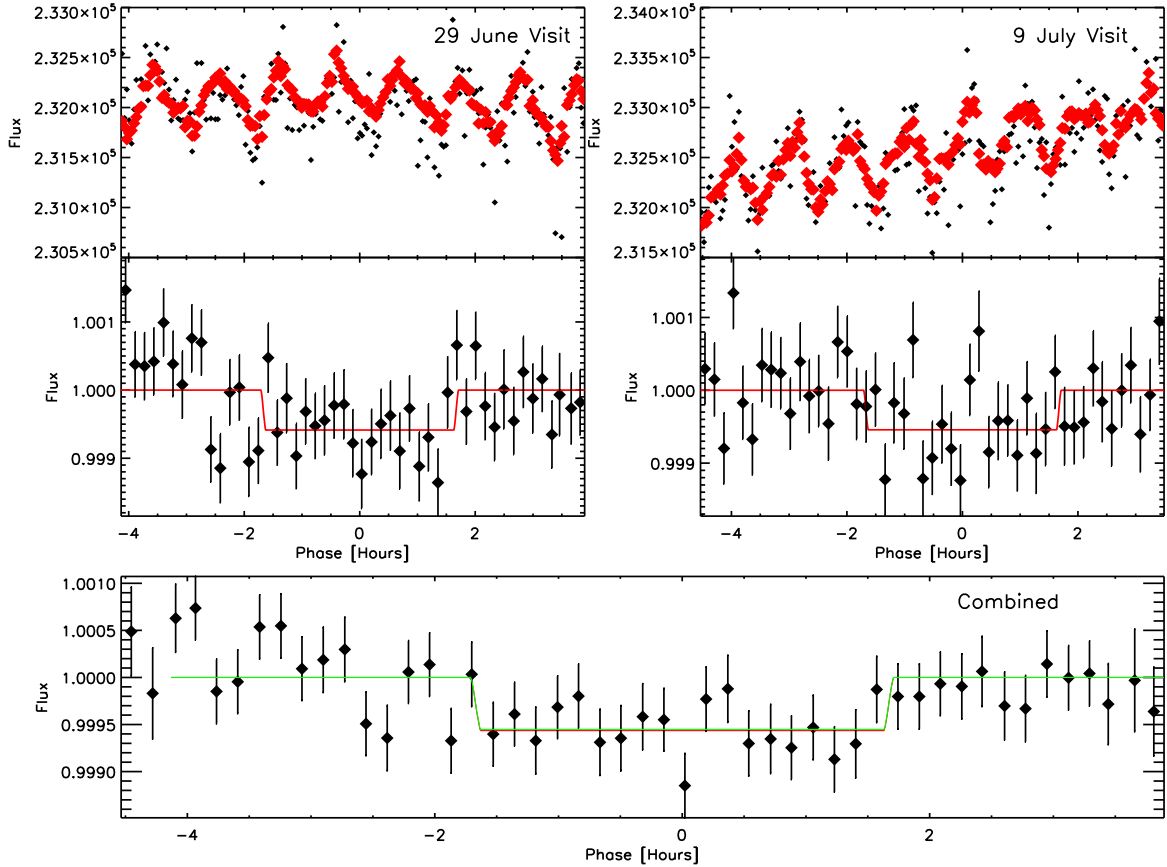} 
 \caption{Both transits of KOI-084 gathered with Warm {\it Spitzer} at 4.5 $\mu$m. The top panels show the raw flux, binned by a factor of 4, with the intrapixel sensitivity variation (obtained with the weighted sensitivity function, as described in the text) overplotted in red. The middle panels show the individual transits with this intrapixel sensitivity removed, binned by a factor of 16, with the best models overplotted. The bottom panel shows the combined transit light curves gathered with \spitzer. The best-fit transit model with depth derived from the \spitzer\ observations is shown in red, while the \kepler\ transit model (corrected for limb darkening) is shown in green. The \spitzer\ and \kepler\ transit depths are in excellent agreement.}
  \label{fig:spitzer}
\end{center}
\end{figure}

We also treated the light curve with the weighted sensitivity function used in \cite{Ballard10}, which proved in that work to produce a time series with lower RMS residuals. For this procedure, we do not assume any {\it a priori} functional form for the intrapixel sensitivity; rather, we perform a weighted sum over neighboring points for each flux measurement, and use this sum to correct each flux measurement individually.  In this way, we build up a map of the intrapixel sensitivity point-by-point. We use same widths, $\sigma_{x}$=0.017 pixels and $\sigma_{y}$=0.0043, for the weighting function (which is a Gaussian in $x$ and $y$) as we used in \cite{Ballard10b}. We therefore have only one free parameter in this case, which is $R_{p}/R_{\star}$. We correct each observation using all other flux measurements, but we did not bin the data, as was done in \cite{Ballard10b}. With this treatment, we also use the residual permutation method to fit the transit depth at each residual permutation. We find $R_{p}/R_{\star}$=0.0242$\pm$0.0032 for the 29 June visit and $R_{p}/R_{\star}$=0.0233$\pm$0.0033 for the 9 July visit. These errors are larger by 20\% and 30\%, respectively, as compared to the MCMC-derived errors. The out-of-transit RMS of the light curve is slightly lower in both cases using the weighted sensitivity function treatment, so in this case we defer to the weighting-function-derived values for $R_{p}/R_{\star}$. Combining the two measurements, we find a radius ratio $R_{p}/R_{\star}$ at 4.5 $\mu$m of 0.0238$\pm$0.0023, which translated to a transit depth of $547^{+113}_{-110}$ ppm. This value is in excellent agreement with the depth in the \kepler\ bandpass of 567$\pm$6 ppm (corrected for limb-darkening). In Figure \ref{fig:spitzer}, we show the combined and binned \spitzer\ light curve, with the best-fit transit model derived from the \spitzer\ observations and the best-fit \kepler\ transit model (corrected for limb-darkening) overplotted. We comment further on the types of blends we rule out with \spitzer\ (and their complementarity with blends ruled out by \blender) in the following section.

\subsection{\blender\ Analysis}

In the absence of a radial velocity confirmation and mass measurement of the planet Kepler-19b, we instead investigate the likelihood that the transit signal is a false positive. Possible false positive scenarios involve another eclipsing system lying within the same photometric aperture as the \kepler\ target star. This binary system could comprise two stars, or a star and a gas giant planet, and could be physically associated or unassociated with the host star (in the foreground or background). In a false positive scenario, the presence of the \kepler\ target star dilutes the depth of the transit to appear planetary (or attributable to a smaller planet, if the binary system comprises a star and a Jupiter-size planet). We employ the \blender\ software package \citep{Torres04, Torres11}, which produces synthetic light curves corresponding to eclipsing binary blend scenarios and attempts to replicate the detailed shape of the \kepler\ transit light curve. The \blender\ technique has been applied previously toward validating three \kepler\ planets: Kepler-9d \citep{Torres11}, Kepler-11f \citep{Lissauer11}, and Kepler-10c \citep{Fressin11}.  Model blend light curves are compared with the \kepler\ photometry in a $\chi^{2}$ sense, with models considered poor fits accordingly deemed unlikely to explain the transit. By exploring the parameter space of mass, impact parameter, orbital eccentricity, and distance from the host star, \blender\ amasses knowledge of the range of possible blends that are consistent with the shape of the transit, and the range of blends that are inconsistent. The \blender\ nomenclature defines the objects within the binary to be the ``secondary'' and ``tertiary'', while the \kepler\ target star is defined to be the ``primary''. In the case of hierarchical triples, \blender\ uses the best isochrone parameters for the \kepler\ target star (derived from SME) as input constraints to the secondary and tertiary stars: these stars are assumed to have formed concurrently and are therefore assumed to be the same age. When the secondary and tertiary are physically unassociated with the \kepler\ target star, \blender\ assumes an age of 3 Gyr (a representative age for the field, per \citealt{Torres11}) and a solar metallicity to model the putative binary system. 

In order to validate Kepler-19b as a planet, we evaluate the probability of false positive scenarios allowable by \blender\ and compare these probabilities to that of the authentic 2.2 $R_{\oplus}$ planet hypothesis. First, we address the probability of a physically unassociated binary in the foreground or background of the target star. In the case of Kepler-19b, all false positive scenarios consisting of a background or foreground eclipsing binary (comprising two stars) are ruled out at the 10$\sigma$ level from the shape of the transit alone. That is, the best planet model furnishes a solution that is a 10$\sigma$ improvement over the best blend model in this case. This result is attributable to the sharp ingress and egress of the transit light curve, which is not well reproduced by blend models involving a binary system consisting of two stars. We therefore confidently conclude that this scenario cannot replicate the \kepler\ transit signal. 

\begin{figure}[h!]
\begin{center}
 \includegraphics[height=4in]{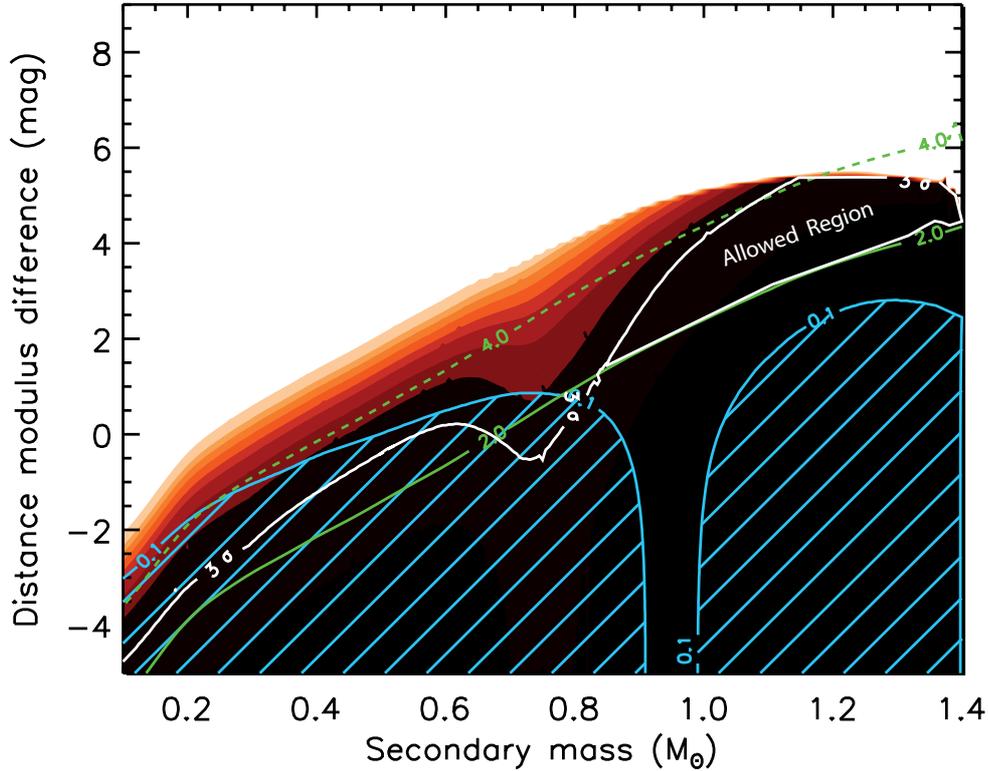} 
 \caption{\blender\ $\chi^{2}$ goodness-of-fit contours corresponding to blend models with background/foreground (physically unassociated) secondary star and planetary tertiary, as a function of distance modulus and mass of the secondary. Viable blend scenarios lie below the 3$\sigma$ contour, depicted with a heavy white line. The blue cross-hatching shows the region of parameter space for which the blend model is the wrong color to be consistent with the measured 2MASS and Sloan $r$ colors of the star, while models that lie below the solid green line have a small enough magnitude difference that the secondary would have been detected in the spectrum of the star ($\Delta m<2.0$). The dashed green line shows location of the $\Delta m$ contrast limit corresponding to the faintest allowable blend. The remaining allowed parameter space is denoted ``Allowed Region.''}
  \label{fig:blender1}
\end{center}
\end{figure}

For scenarios consisting of a foreground/background star orbited by a larger (Jupiter-size) planet, there exists a region of parameter space in which an eclipsing binary model provides a comparably good fit, as compared to the single star and planet model. The \blender\ constraints are represented by contours of equal goodness of fit in Figure \ref{fig:blender1}. The 3$\sigma$ contour is shown with a heavy white line, and blend scenarios under this curve are considered viable. In additional to the goodness-of-fit of the blend models to the \kepler\ light curve, there are regions of parameter space that are disallowed by the color of the star (as measured by 2MASS and in Sloan $r$ band), as well as the spectrum (in which a secondary star within a certain magnitude range would be apparent). These constraints are depicted in Figure \ref{fig:blender1} as blue cross-hatching (within which region blends are disallowed) and a solid green line (below which an additional star would have appeared in the spectrum). We comment briefly on this spectroscopy constraint, which we measured by injecting additional stellar spectra (in this case, solar-type) into the spectrum of Kepler-19 at varying brightnesses and relative velocities, and then determining the limits on detectability via a cross-correlation of the spectrum with a template. We determine that any star within 10\% the flux of Kepler-19 would be detectable in the cross-correlation function down to relative velocities of 5 km s$^{-1}$, which translates conservatively to a $\Delta m<2$ constraint. A velocity variation of $<$5 km s$^{-1}$ would be unlikely for a random unassociated background star, and we comment on the hierarchical triple case further below. Furthermore, the angular separation of this star and planet system must also be sufficiently small as to be undetectable by adaptive optics imaging for which limits are given in Section \ref{sec:ao}. While an unassociated binary might be resolvable by adaptive optics, for a hierarchical triple the possibility of an unresolved companion remains.

We proceed to evaluate the frequency of the remaining allowable blends as follows: We estimate {\it a priori} the frequency of stars in the background or foreground of the target star. We evaluate a theoretical number density of neighboring stars, per square degree, using Galactic structure models given by \cite{Robin03}. We record this number density in half-magnitude bins (shown in Table \ref{tbl:blender}), and show both the number density of stars and their allowed mass values in Columns~3 and 4, based on the constraints constraints imposed by the  \blender\ contours, additional color constraints, and brightness constraints. For the magnitude bins in which no blend furnishes a solution within 3$\sigma$ of the planetary model, we have left these columns blank. The maximum angular separation at which these stars might have remained undetected in adaptive optics imaging (the limits are provided in Section \ref{sec:ao}) is listed in Column~5. The number of stars in each magnitude bin is then listed in Column~6. In order to evaluate the transiting planet prior, we rely on the reported \kepler\ planet candidate sample to date, presented in \cite{Borucki11}. While the majority of these candidates have not yet been confirmed to be planets, the false positive rate is expected to be quite small (as reported by \citealt{Morton11}) and so will not substantially change our results. We also assume that the sample presented in the \cite{Borucki11} candidate list is complete. 

\begin{figure}[h!]
\begin{center}
 \includegraphics[height=4in]{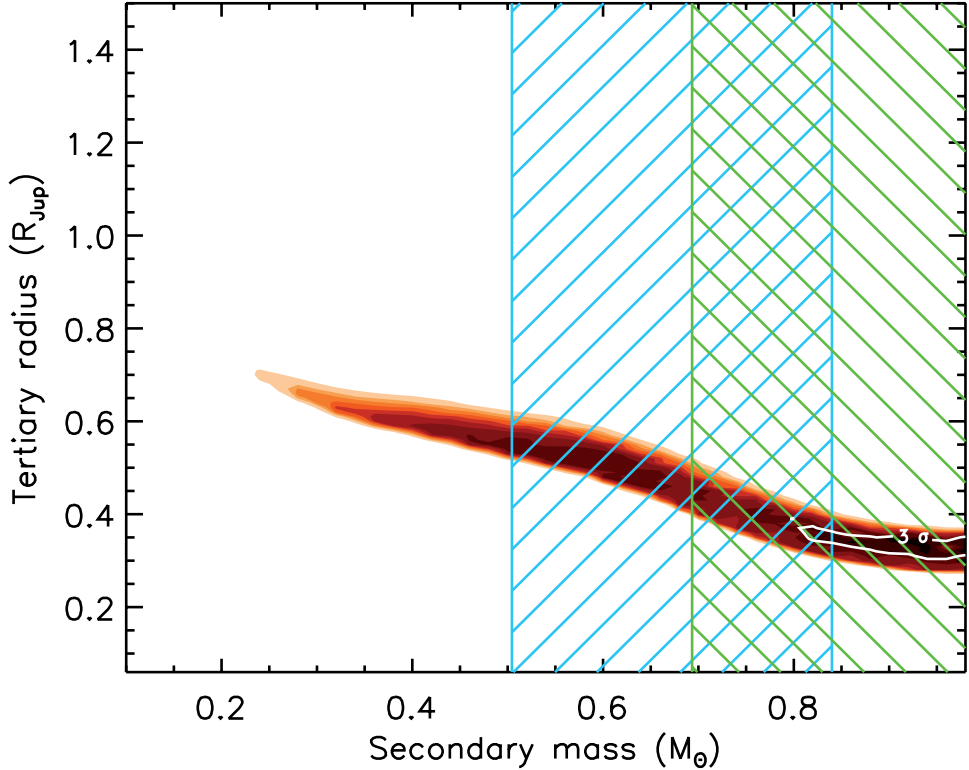} 
 \caption{\blender\ $\chi^{2}$ goodness-of-fit contours corresponding to hierarchical triple blend models (with physically associated secondary star and planetary tertiary). As in Figure \ref{fig:blender1}, viable blend scenarios lie within the 3$\sigma$ contour, depicted with a heavy white line. The color-coding is similar to the previous figure, with blends of the wrong color depicted in blue cross-hatching, and blends that are ruled out spectroscopically depicted in green cross-hatching.}
  \label{fig:blender2}
\end{center}
\end{figure}

The second feasible blend scenario is an additional star and transiting Jupiter-size planet, which are physically associated with the \kepler\ target star. We present the results of \blender\ for this case in Figure \ref{fig:blender2}. While there exists a range of hierarchical triples whose light curve shape is consistent with the \kepler\ transit (depicted by the $\chi^{2}$ contours), these are all ruled out by either the color constraint on blends (shown in blue cross-hatching) or the brightness constraint on blends from the spectrum (shown in green cross-hatching). There exists the remaining possibility of a true twin to the target star: a star that has an identical color, and whose position is either at a distance $>$ 20 AU (at which position the predicted radial velocity is equal to 5 km s$^{-1}$) or whose tangential velocity is $<$5 km s$^{-1}$ relative to the target star during the time of our observations. Even if this scenario were to be true, the planet's inferred radius would only be greater by a factor of $\sqrt{2}$. Therefore, we conclude that the only possible blends belong to the unassociated planet and star scenario.

Combining the probabilities associated with all background or foreground star and planet pairs, we find a total blend frequency of 1.08$\times10^{-7}$. This frequency corresponds to the likelihood of a blend that is capable of producing a transit light curve that is no worse ($<3\sigma$) than the best model corresponding to a transiting planet around a single star within the aperture. 

Next, we estimate the {\it a priori} frequency of a true planet with the characteristics implied by the \kepler\ transit light curve. Using the planetary radius range of 2.209$\pm$0.048 $R_{\oplus}$, we identify 119 planet candidates from the \cite{Borucki11} catalog with sizes that are within 3$\sigma$ of this measured value. The planet prior is equal to 7.6$\times10^{-4}$ (119 divided by the total number of \kepler\ targets, 156,453), which is more than 3 orders of magnitudes larger than the blend frequency. We therefore find that the planetary scenario corresponding to a 2.2 $R_{\oplus}$ planet is 7000 times as likely as the blend scenario, and conclude with very high confidence that the transit signal is due to a planet, Kepler-19b.

For comparison, the constraint from \spitzer\ in this case provides an independent means of ruling out a subset of blends which are also ruled out by \blender. If we impose the constraint that a putative additional star may not produce a transit depth at 4.5 $\mu$m which is 3$\sigma$ deeper than measured, such a star cannot be less massive than 0.7 $M_{\odot}$ (otherwise, the additional star would be so red as to produce a significantly deeper transit depth in the near-infrared). As shown in Figures \ref{fig:blender1} and \ref{fig:blender2}, \blender\ independently rules out blends consisting of a star in this mass range from the shape of the transit light curve.

\begin{deluxetable}{ccccccc}
\tablenum{4}
\tabletypesize{\scriptsize}
\tablewidth{0pc}

\tablecaption{Blend frequency estimate for KOI-084.01. \label{tab:blendfreq}}
\label{tbl:blender}
\tablehead{
& &
\multicolumn{5}{c}{Blends Involving Planetary Tertiaries} \\[+1.5ex]
\cline{3-7} \\ [-1.5ex]
\colhead{$K\!p$ Range} &
\colhead{$\Delta K\!p$} &
\colhead{Stellar} &
\colhead{Stellar Density} &
\colhead{$\rho_{\rm max}$} &
\colhead{Stars} &
\colhead{Transiting Planets} \\
\colhead{(mag)} &
\colhead{(mag)} &
\colhead{Mass Range} &
\colhead{(per sq.\ deg)} &
\colhead{(\arcsec)} &
\colhead{($\times 10^{-6}$)} &
\colhead{0.36--2.00\,$R_{\rm Jup}$, $f_{\rm planet}=0.20$\%} \\
\colhead{} &
\colhead{} &
\colhead{($M_{\odot}$)} &
\colhead{} &
\colhead{} &
\colhead{} &
\colhead{($\times 10^{-6}$)} \\
\colhead{(1)} &
\colhead{(2)} &
\colhead{(3)} &
\colhead{(4)} &
\colhead{(5)} &
\colhead{(6)} &
\colhead{(7)}
}
\startdata
11.9--12.4  &  0.5 & \nodata   & \nodata&\nodata & \nodata& \nodata \\
12.4--12.9  &  1.0 & \nodata   & \nodata&\nodata & \nodata& \nodata \\
12.9--13.4  &  1.5 & \nodata   & \nodata&\nodata & \nodata  & \nodata \\
13.4--13.9  &  2.0 & \nodata & \nodata   & \nodata  & \nodata  & \nodata \\
13.9--14.4  &  2.5 & 0.88--1.40  & 444   &  0.22  &  5.21  & 0.011 \\
14.4--14.9  &  3.0 & 0.91--1.40  & 505   &  0.29  &  10.3  & 0.021 \\
14.9--15.4  &  3.5 & 0.95--1.40  & 436   &  0.38  &  15.3  & 0.031 \\
15.4--15.9  &  4.0 & 1.00--1.30  & 327   &  0.53  &  22.3  & 0.045 \\
15.9--16.4  &  4.5 & \nodata     & \nodata&\nodata & \nodata& \nodata \\
16.4--16.9  &  5.0 & \nodata     & \nodata&\nodata & \nodata& \nodata \\
16.9--17.4  &  5.5 & \nodata     & \nodata&\nodata & \nodata& \nodata \\
17.4--17.9  &  6.0 & \nodata     & \nodata&\nodata & \nodata& \nodata \\
17.9--18.4  &  6.5 & \nodata     & \nodata&\nodata & \nodata& \nodata \\
18.4--18.9  &  7.0 & \nodata     & \nodata&\nodata & \nodata& \nodata \\
18.9--19.4  &  7.5 & \nodata     & \nodata&\nodata & \nodata& \nodata \\
19.4--19.9  &  8.0 & \nodata     & \nodata&\nodata & \nodata& \nodata \\
19.9--20.4  &  8.5 & \nodata     & \nodata&\nodata & \nodata& \nodata \\
\noalign{\vskip 6pt}
\multicolumn{2}{c}{Totals} & & 1712 &  & 53.11 & 0.108 \\
\noalign{\vskip 4pt}
\hline
\noalign{\vskip 4pt}
\noalign{\vskip 4pt}
\hline
\noalign{\vskip 4pt}
\multicolumn{7}{c}{Total frequency (BF) = $(0.108) \times 10^{-6} = 1.08 \times 10^{-7}$} \\
\enddata

\tablecomments{Magnitude bins with no entries correspond to brightness
ranges in which \blender\ excludes all blends.}

\end{deluxetable}
\clearpage

\section{Discussion and Conclusions}

\subsection{Interpretation of Transit Timing Variations}

Our analysis of the transit timing variation of Kepler-19b comprises two sections. In Section~\ref{sec:planetproof}, we argue that there must exist a second planet in the Kepler-19 system, since the TTVs cannot originate from astrophysical effects or a stellar mass perturber. Then, in Section~\ref{sec:planetid}, we describe the dynamical properties of planetary perturbers that could account for the observed TTV pattern of planet b.

\subsubsection{Demonstration of Planethood of Perturber} \label{sec:planetproof}
In Figure 4 we presented a strongly detected and nearly sinusoidal variation (period $P_{\rm ttv}=316$~days and amplitude $A_{\rm ttv}=5$~min) in the times of transits of planet b.  Here we discuss four potential interpretations of the signal which do not invoke a perturbing planet, and demonstrate that the signal cannot originate from these scenarios. The first two scenarios consider a system of only the star and the transiting planet, while the latter two allow for the presence of a third, non-planetary, body. Because these scenarios are disallowed, the only viable alternative is that the signal is a dynamical variation caused by a second planet (planet c), which we discuss in the next subsection. 

First, we consider the possibility that the signal is due to stellar activity.  The most plausible candidate for dynamical interaction with the star causing the TTV signal is the \cite{Applegate92} effect from the eclipsing binary literature, which \cite{Watson10} recently applied to exoplanets.  In this mechanism, the star undergoes a magnetic cycle, which varies the rotational bulge's gravitational pull on the planet, slightly varying its orbital period.  For this effect to produce the observed TTV signal, the magnetic cycle would need to have an exceptionally short period of $316$~days.  This is problematic because the stellar activity in the spectra is low, as we describe in Section 4.2, suggesting a magnetically inactive star with a long spin period. For magnetic dynamos typical of solar-type stars like Kepler-19, \cite{Watson10} calculate TTV variations of less than $1$ second over timescales of several years, much too small to explain our data.  Apart from this dynamical interaction, activity could cause apparent TTV via the planet transiting over starspots \citep{SilvaValio08,Alonso08}. For instance, \cite{Knutson11} recently found transit timing deviations for GJ~436b that are greater in optical photometry, where spots are pronounced, than in near-infrared photometry, where spots are relatively muted.  However, in our case, even with excellent \kepler\ photometry, no spots are detected, either in out-of-transit stellar modulation or in excess residuals during transit due to spot crossings.  

Second, the signal could be due to rotation of the planetary orbit's apsidal line about the star \citep{Heyl07}. The eccentricity need not be large ($e_b \simeq 2 \pi A_{\rm ttv} / P_b = 0.0023$), but the apse must be precessing very quickly to be consistent with the TTV period ($P_{\rm ttv}=316$~days).  Using expressions for realistic apsidal motion rates \citep[section 3.1.1]{Fabrycky10}, we demonstrate that the periods associated with possible precession mechanisms are too long by several orders of magnitude, as follows. General relativistic precession has a period of $7.9 \times 10^4$~yr.  A star made oblate by rotation, alternatively, would cause precession with a period of $7 \times 10^6$~yr~$\times(P_{\rm rot} / 10~{\rm days})^2$, if the star has an apsidal motion constant of $k_{\rm L}/2 = 0.02$ \citep{Claret92}.  Finally, tidal distortion of the planet would cause precession with a period of $\sim 10^8$~yr, if the planet has a Love number of $k_{\rm L} = 0.3$ \citep{Mardling04}.   We conclude that general relativity dominates the putative precession rate, but it is inconsistent by orders of magnitude with $P_{\rm ttv}$ and is probably undetectable \citep{Ragozzine09}.

Third, the signal could originate from light time delay, owing to reflex motion of the system as it is orbited by a third body with an orbital period of $P_2 = P_{\rm ttv}\simeq316$~days.  It is conceivable that this putative body moves the barycenter of the Kepler-19 / Kepler-19 b system by $\pm 5$~light minutes, causing a time-variable light-time delay (e.g., \citealt{Irwin52,Sybilski10}).  In creating such a large displacement in the radial direction, a radial velocity signal -- its derivative -- would also be created.  The semi-amplitude would be $2 \pi (0.6 {\rm AU})/(316 {\rm days}) \simeq 17$~km s$^{-1}$, ruled out by orders of magnitude given our radial velocity measurements.  Moreover, to induce this motion, the additional body must have a mass of at least 1.0$M_\odot$, and would likely impart a second set of lines on the spectra, which are not detected. 
 
Finally, the signal could be dynamical, owing to perturbations from another body in the system such as a second star or a brown dwarf.  The orbital period of such a body cannot be too long; otherwise, the gravitational potential it induces on planet b would result in a period longer than the observed $P_{\rm ttv}$.  In fact, the longest period the body could have is $2 P_{\rm ttv}$ \citep{Borkovits03}, as the dominant part of the perturbation of a distant body is its ``tidal'' term, which has a frequency of twice the body's orbital frequency\footnote{Planet b would also speed up and down twice per orbit due to a static tidal term, but this is not observable, as transits occur only once per orbit.}.  The lack of large radial velocity variation means that additional bodies must have masses in the planetary regime (a case handled in the next subsection) on nearly any orbit with $P_2 \leq 2 P_{ttv}$.  An exception is for nearly face-on orbits with a small component of radial motion.  To our knowledge, there is only one such configuration that could explain the timing data, as follows.  A circular orbit of period $2 P_{ttv}$, mutually perpendicular to the orbit of planet b, causes a timing signature \citep[eq. 46]{Borkovits03} :
\begin{equation}
O-C_b \approx \frac{3}{8 \pi} \frac{M_2}{M_\star + M_b + M_2} \frac{P_b^2}{P_2} \sin 2 f_2 
\end{equation}
where $M_2$ is the perturber's mass and $f_2$ is its true anomaly measured with respect to the plane of planet b's orbit.  The perturbing orbit must be nearly circular in the case of Kepler-19, as we have measured two full $O-C$ cycles and found them to be nearly identical, rather than different either in amplitude or phase as would result from an eccentric perturber (e.g., run A13 of Figure 2 in \citealt{Borkovits03}).  To fit $A_{ttv}$, we have $M_2 \approx 0.25 M_\odot$.  The radial velocity limit (at 99.73\% confidence) on circular orbits at $P_2 = 2 P_{ttv}$ is $K<22$~m~s$^{-1}$, so the orbit would need to be inclined by $i_2<0.2^\circ$ to the plane of the sky.  Given an isotropic prior, this configuration has a probability of $1-\cos i_2 = 6.0 \times 10^{-6}$, i.e. it is too finely tuned toward face-on to be plausible.

Having thus demonstrated that the TTV signal cannot be created by any known mechanism other than a second planet, we interpret the transit timing effect of planet b to be due to a second planet in the system, which we call Kepler-19c.
 
\subsubsection{ Possible orbits of planet c } \label{sec:planetid}
 
In this subsection we discuss the possible orbits of planet c, consistent with the TTV data for planet b.  We take as constraints a sinusoidal transit timing signal as well  as radial velocity upper limit of $M_c < 0.44 M_{\rm Jup} \times (P_c/P_b)^{1/3}$ (99.73\% confidence limit on a second orbit of arbitrary eccentricity).  This assumes $\sin i_c > 0.5$, as we consider a mutual inclination $i_m>60^\circ$ to be physically implausible\footnote{A large mutual inclination would likely drive large-amplitude eccentricity cycles \citep{Kozai62} in the innermost planet, which would in turn trigger rapid tidal orbital decay \citep{Fabrycky07,Mardling10}.}. These planetary scenarios fall broadly into five categories: orbits with the period of the TTV signal, resonant perturbers, orbits near first-order mean-motion resonances, orbits near higher-order resonances (in which category we assign an upper limit on the mass of the perturber), and satellite scenarios. The latter three are favored, under circumstances that we elucidate below.

Our first consideration is the possibility that planet c could have a period $P_c = P_{\rm ttv} = 316$~days and a large eccentricity, such that it produces a time-variable tidal force on the Kepler-19 / Kepler-19 b pair, accounting for the TTVs \citep{Borkovits03,Agol05,Borkovits11}.  However, radial velocity constraints require its mass is $M_c\lesssim 1.4 M_{\rm Jupiter}$, so to generate a TTV signal with amplitude $A_{\rm ttv}=5$~min, its eccentricity would need to be $>0.99$ \citep{Agol05}.  This would be unstable with respect to planet b because the orbits would cross.  Moreover, a smaller amplitude could be generated at $0.5<e_c<0.9$, but the signal would have a saw-toothed shape, which is inconsistent with the measured transit times.  Therefore, any putative planet with period $316$~days that is massive enough to create the TTV signal would be inconsistent with the observed radial velocities.

Another possibility is that the TTV curve of b is driven by a resonant perturber.  In this case, the amplitude can be substantial, even for a low-mass perturber \citep{Agol05},
\begin{equation}
\delta t \approx \frac{P_b}{p} \frac{M_c}{M_b+M_c},
\end{equation}
where $p$ refers to the period ratio of the transiting planet to the perturbing planet $p/q$, $\mu \equiv {\rm max}(M_b,M_c) / M_\star$, and $\delta t$ is the amplitude of the TTV signal. The libration (and TTV) period is of order $P_{\rm lib} \simeq e^{-1/2} \mu^{-1/2} P / p$ \citep{Agol05}.  This consideration determines whether such a resonance is a possible site for planet c, as the mass of c must be large enough so that this libration period equals the rather short TTV period.  For low eccentricities ($e_b, e_c \lesssim 0.1$), the mass of the perturbing planet must be $\sim 1$ Jupiter mass, which is ruled out by the radial velocities.  For high eccentricities, lower masses are allowed, but then the region of stability becomes more constrained, such that the proposed system would need to be finely tuned. Orbits near the Lagrange points of b (the 1:1 resonance; \citealt{Ford07}) fail for the same reason: $P_{\rm lib} \simeq P_b \sqrt{ 27/4 \times M_\star / (M_b + M_c) }$ \citep{Ford07}, which means $M_c \simeq 3 M_{\rm Jup}$, in violation of the radial velocity constraint. For these reasons, we do not favor resonant orbits for planet c. 

Next, we consider perturbers near period commensurabilities.  If two planets are near to, but offset from, a period commensurability, they can generate a large TTV signal \citep{Agol05,Holman05}, as in the interaction between planets Kepler-11 b and c \citep{Lissauer11}.  In this scenario, the orbital frequency ratio of the known planet to the perturbing planet, $P_b / P_c$, is slightly offset from a ratio of integers, $p / q$.  Then, as long as the planets' eccentricities are moderate, the transit timing signature will have a dominant frequency equal to:
\begin{equation}
1/P_{\rm ttv} = | p / P_b - q / P_c |.  \label{fttv}
\end{equation}
The absolute value sign above means we can postulate a perturber on one side or the other of each resonance, that is responsible for the TTV signal. The strength of the resonance must be finely tuned, so that a 5 minute amplitude signal is possible, yet the radial velocity amplitude also lies within the observed upper limit.  If the resonance is spaced too closely to the transiting planet, the planets will generate timing variations on the conjunction timescale \citep{Nesvorn08,Holman10} which are not observed, or else the two planets will not be stable with respect to one another \citep{Wisdom80}.  

The absolute value sign above means we can postulate a perturber on one side or the other of each resonance, that is responsible for the TTV signal. The resonance must be strong enough that a perturber small enough to fulfill the radial velocity constraints can produce the 5 minute amplitude timing signal.  However, if the resonance is spaced too closely to the transiting planet, the planets will generate timing variations on the conjunction timescale \citep{Nesvorn08,Holman10} which are not observed, or else the two planets will not be stable with respect to one another \citep{Wisdom80}.
 
For first-order resonances in which $q=p \pm 1$, \cite{Agol05} developed an order of magnitude estimate for TTV signals which depends on the fractional offset from the resonance $\epsilon \equiv | 1- q P_b / (p P_c) |$ and the mass ratio $\mu \equiv {\rm max}(M_b,M_c) / M_\star$:
\begin{eqnarray}
\delta t_b = \left\{
\begin{array}{cl}
(M_c/M_\star) \epsilon^{-1} P_b, &\qquad \epsilon \geq \mu^{1/2} \\
{\rm min}(M_c/M_b, 1) \mu^2 \epsilon^{-3} P_b, &\qquad \epsilon \leq \mu^{1/2}.
\end{array}
\right\} \label{eqn:ttvamp}
\end{eqnarray}
In the current case, presuming planetary masses of $\lesssim10 M_{\oplus}$, the upper expression of Equation \ref{eqn:ttvamp} holds for $p \lesssim 5$. In particular, for exterior first-order resonances $q$:$p$ = (2:1, 3:2, 4:3), to produce $A_{\rm ttv} = 5$~min we have $M_{c} \sim (4, 2, 1)M_{\oplus}$. Of course, these values are only good to order-of-magnitude, but they demonstrate that a reasonable planetary mass just offset from first-order resonances can indeed cause the observed TTV signal. 

A planet near a second-order (e.g., 3:1, 5:3) or higher-order (e.g., 4:1, 5:1) resonance can also be responsible for the TTV signal.  However, in these cases the strengths of the resonances are smaller, and they depend on a higher power of the eccentricities.  Therefore, planetary-mass perturbers that satisfy the radial velocity constraints might need to have substantial eccentricities to match $A_{\rm ttv}$. For $n:1$ resonances of exterior perturbers with $n\gg1$, $e_c$ must be large, and the timing signal would be a series of constant-period segments with kicks at the outer body's periastron passage.  The scaling relation of the radial velocity limit quoted above breaks down at high eccentricity, as periaston passages can appear as spikes which fall in data gaps, but we still wish to limit $M_c$.  We therefore set 99.73\% confidence limits on $M_c \sin i_c$ by (1) introducing a 4.8 $M_{\oplus}$, circular and edge-on planet b, and (2) sampling the orbit of planet c on a grid with $P_c$ drawn from equation~\ref{fttv} for all $n:1$ resonances with $n=3$ to $16$ (28 cases); $e_c$ drawn from $0.1$, $0.2$ ... until the orbit crossed with planet b's; $\omega$ drawn from $0^\circ$, $45^\circ$, ... $315^\circ$; $T_{0,c}$ drawn from 10 values uniformly spaced between BJD $2455200$ and $2455200+P_c$; and $M_c$ spaced logarithmically by 0.25 dex from $0.32M_{\rm Jup}$ to $18 M_{\rm Jup}$.  The total grid sampled 150120 trials, and not one of the cases at or above $5.6M_{\rm Jup}$ fit the radial velocities (allowing for a constant offset).  Apparently, although a high-mass perturber can be fine-tuned to not induce large radial velocity at the times of the data, its mass is limited even at arbitrary eccentricity.  For concreteness, let us describe the end-member of this set of n:1 resonances.  It has $P_c = 153.2 $ days $ \approx P_{ttv}/2$, and the~$O-C$~signal would be a zig-zag, only marginally consistent with the apparently sinusoidal shape.  At each periastron passage of planet c, $P_b$ would need to change by $0.6$~minutes to match $A_{ttv}$.  A consistent set of parameters according to \cite[eqn. 2]{Holman05} is 1~$M_{\rm Jupiter}$ and $e_c \gtrsim 0.8$.  The periastron of planet c would thus be at $\lesssim 1.3 a_b$, only marginally stable with respect to the inner planet.  This scenario is also barely allowable according to the radial velocities.  Therefore we set an upper limit of $P_c \lesssim 160$~days.  Again, this corresponds to $M_c \lesssim 6 M_{\rm Jupiter}$, according to our grid search.

The final possibility that we consider is a satellite orbit, which could also cause TTVs (e.g., \citealt{Kipping09}).  The amplitude of 5 minutes translates to a displacement along the orbit of $2.1R_p$. A prograde orbit for such a satellite would lie between $R_p$ and $\sim 7 R_p$ ($0.4$ of the Hill sphere, if $M_b \simeq 5 M_{\rm Earth}$) to be stable \citep{Domingos06}.  Since the putative satellite lies so close to the planet in this case, its mass would have to be $\gtrsim 0.6 M_b$ cause the TTVs observed for the transiting planet.  Therefore, it would probably be big enough to be seen in transit.  We examined each transit by eye, to see if any deviated significantly from the single-planet model, as mutual events of the co-orbiting planets would cause shallower transits \citep{Szabo06, Simon07, Ragozzine10}, but we found no features of interest. Furthermore, in this scenario, the b-c mutual orbital period would need to be near-resonant with the pair's orbital period around the star, so that the TTV signal aliases to the long $P_{\rm ttv}=316$~day signal.  We find this scenario unlikely. 

A retrograde orbit for a satellite could be stable to much larger distances, even beyond the Hill sphere \citep{Jackson13,Shen08}.  In fact, planet c could follow an independent Keplerian orbit which resonates with planet b, keeping their periods and orbital phases the same \citep{Laughlin02}.  The allowable $M_c$ could be much lower in this case, down to of order $0.1 M_{\rm Earth}$ if the eccentricity is $e_{c} \sim 0.1$.  The TTV could be caused by either the resonant libration or by the two orbits precessing together at a swift rate.  If the latter, then as in Section~\ref{sec:planetproof}, $e_b$ would be $\sim 0.0023$; unlike in that section, an apsidal motion period of 316 days is plausible because of the proximity of the perturber.

\begin{figure}[h!]
\begin{center}
 \includegraphics[height=4in]{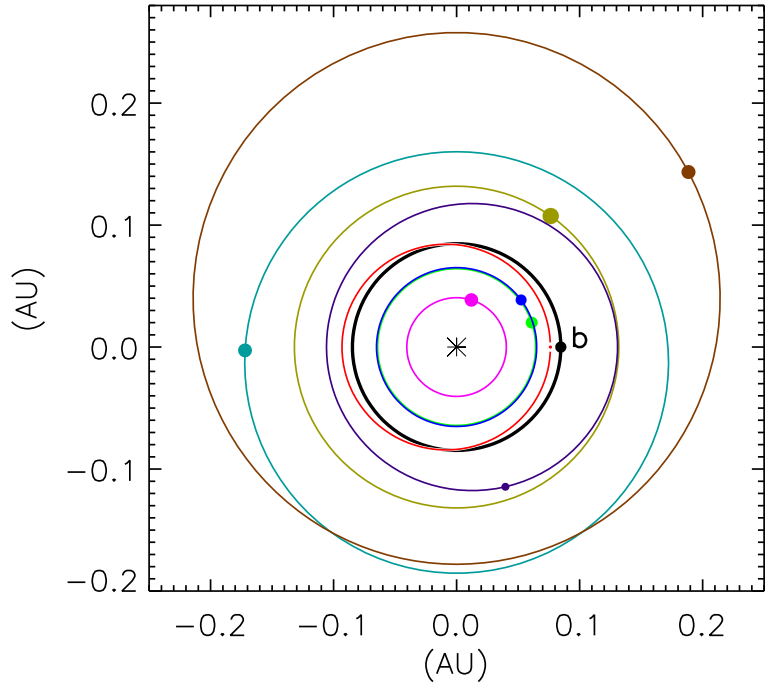} 
\caption{ Possible orbits for Kepler-19c.   Orbits near first-order mean-motion resonances (eq.~\ref{fttv}) may fit the TTV signal and RV constraints even on circular orbits; shown are $P_c = 6.129$~days (\emph{bright green}) and $6.256$~days (\emph{blue}), flanking the interior 2:3 resonance, and $P_c=18.033$~days (\emph{olive green}), next to the exterior 2:1 resonance.  Orbits near higher-order resonances likely need eccentricity to produce the TTV signal with masses low enough to satisfy the radial velocities; shown are $P_c=3.065$~days (\emph{pink}) near the interior 1:3 resonance, $P_c=15.326$~days (\emph{purple}) near the exterior 5:3 resonance, $P_c=27.036$~days (\emph{aqua blue}) near the exterior 3:1 resonance, and $P_c=38.310$~days (\emph{brown}) near the exterior 4:1 resonance.  Finally, a co-orbital planet (or distant retrograde satellite) is shown at $P_c=9.287$~days (\emph{red}) -- in such an orbit very small masses are possible, so the dot representing the planet is drawn small.  Other possible orbits are within mean motion resonances (including the 1:1 at Lagrange points), or a prograde satellite, but these are disfavored (see text).}
\label{fig:planetc}
\end{center}
\end{figure}

We compiled examples of our favored orbits for planet c into Figure~\ref{fig:planetc}.  Foremost are orbits nearby first-order mean motion resonances, which can fit the TTVs with masses which comfortably obey the RV constraints, and have near-circular orbits.  Second, we consider orbits nearby higher order resonances to be possible, particularly if eccentricities are non-zero.  Third, a retrograde satellite is a possibility, but we recognize this option as an exotic one.  There is clearly profound degeneracy of interpretation.  Such degeneracies are intrinsic to the TTV method of planet discovery, in the case that the signal is well-characterized by a single sinusoid \citep{Nesvorn08} and the radial velocity data cannot pinpoint the perturber \citep{Meschiari10}.

\subsection{Constraints on Transits of Perturber}
The \kepler\ team searches for transit signatures using the Transiting Planet Search pipeline module \citep{Jenkins10}. This has already been applied to the Kepler-19 light curve, and it identified transits of Kepler-19b. We analyzed the size of the planet we could have detected in the \kepler\ photometry by injecting signals at random phases and at varying planetary radii and attempting to blindly recover them. While the mass of Kepler-19c is highly uncertain, we can rule out the transits of at least some putative perturbers, as we describe here. The degeneracy of interpretation of the TTV signal means that we cannot address an exhaustive list of potential planets, but we comment here on the representative cases shown in Figure \ref{fig:planetc}. The perturbing planet Kepler-19c may also not transit, in which case its orbit may be significantly misaligned from the transiting planet. Because Kepler-19b resides in a near-equatorial orbit (with $i>$88.62 with 3$\sigma$ confidence), the orbit of Kepler-19c may also be misaligned by at least 0.27$^{\circ}$, or 3.6$\sigma$ in the exterior 2:1 resonance, for example, if it does not transit. In contrast, in the interior 1:2 resonance at 4.57 days, the planet would have to have $i<$85.7 to avoid transit. The planets would have to be misaligned by nearly 3 degrees in that case, if the orbits are circular. 

First, we consider the orbits of 19c in second-order mean-motion or higher-order resonances. The planets depicted in Figure \ref{fig:planetc} range from 1.6 to 13.8 $M_{\oplus}$. We evaluate a minimum physical radius for the 1.6 $M_{\oplus}$ planet, assuming maximum collisional stripping of the planet during formation, from the relationship derived by \cite{Marcus10}. At the maximum possible iron fraction, a 1.6 $M_{\oplus}$ planet would have a radius of 1 $R_{\oplus}$. At a period of 30 days (which range encloses mean-motion resonances up to 3:1 with Kepler-19c) we achieve 95\% completeness at 1.0 $R_{\oplus}$. At 40 days, which includes the example perturber in Figure \ref{fig:planetc} in the 4:1 mean-motion resonance, we would still have detected a 1.0 $R_{\oplus}$ planet at 90\% of phases. 

If the planet Kepler-19c were instead coorbital with 19b, or if it resides in a satellite orbit, its mass could be much smaller, as described in the previous section: this mass could be as smaller as 0.1 $M_{\oplus}$ (equivalent to the mass of Mars) if its eccentricity were equal to 0.1. The models of \cite{Seager07} show that a 0.1 $M_{\oplus}$ planet could be as small as 0.4 $R_{\oplus}$ if it comprised 70\% iron and 30\% silicate by mass (which is plausible, given the maximum iron fractions determined by \citealt{Marcus10}). Its predicted transit depth would be 20 ppm, which might be barely detectable when compared to the error bar on the transit depth of Kepler-19b (with a necessarily similar orbital period) of 6 ppm. However, even if such a planet transited, its detection would be extremely challenging.

 However, though the representative cases (with the exception of the coorbital scenario) depicted in Figure \ref{fig:planetc} would all have been readily detectable, we note that the perturbing planet could easily be smaller than 1 $R_{\oplus}$ in the mean-motion or higher-order resonance cases, in which case it might have eluded detection even in orbits with periods shorter than 30 days. However, we note that we achieve 95\% completeness for 0.7 $R_{\oplus}$-sized planets up to 10 days, so such a world would have to be less massive than 0.7 $M_{\oplus}$ for $P<10$ days, referring again to the maximum iron fraction models derived by \cite{Marcus10}.

\subsection{Search for Secondary Eclipse of Kepler-19b}
If we assume the planet reradiates isotropically the energy it receives from its star, then the equilibrium temperature of Kepler-19b is given by:

\begin{equation}
T_{p}=(1-A_{B})^{1/4}T_{\star}\sqrt{\frac{R_{\star}}{2a}},
 \label{eq:temp}
\end{equation}
 \noindent where $T_{\star}$ is the temperature of the star, $a/R_{\star}$ is the orbital radius to stellar radius ratio, and $A_{B}$ is the albedo of the planet. If we assume a Bond albedo $A_{B}$ of 0.3 and ignore atmospheric effects in order to obtain a rough estimate for the equilibrium temperature, we may employ the MCMC chain of $a/R_{\star}$ (and the corresponding values for $T_{\rm eff}$ of the star identified from the nearest stellar isochrone, per the analysis in Section 3.2) to find the allowable range of planetary temperature. The range that encompasses 68\% of realizations of planetary temperature is $T_{P}=770 \pm 10$ K, rounded to the nearest 10 Kelvin. 

There are two contributing sources to the occultation depth, namely the reflected starlight, and the emitted light from the planet itself. Both of these are dependent upon the unknown albedo of the planet. Assuming again the Bond albedo of 0.3, the expected depth due to reflected light is given by $\delta_{ref}=A_{B}(R_{p}/a)^{2}$; this value lies between 3 and 4 parts in 10$^{7}$. The expected occultation depth due to the emitted light of the planet is given by $\delta_{em}=(R_{p}/R_{\star})^{2}\cdot B_{\lambda}(T_{p})/B_{\lambda}(T_{\star})$. Using a wavelength of 700 nm (in the middle of the \kepler\ bandpass) to estimate $B_{\lambda}(T)$, $\delta_{em}$ is of order $10^{-13}$ and so contributes negligibly to the expected eclipse depth. 

\begin{figure}[h!]
\begin{center}
 \includegraphics[height=4in]{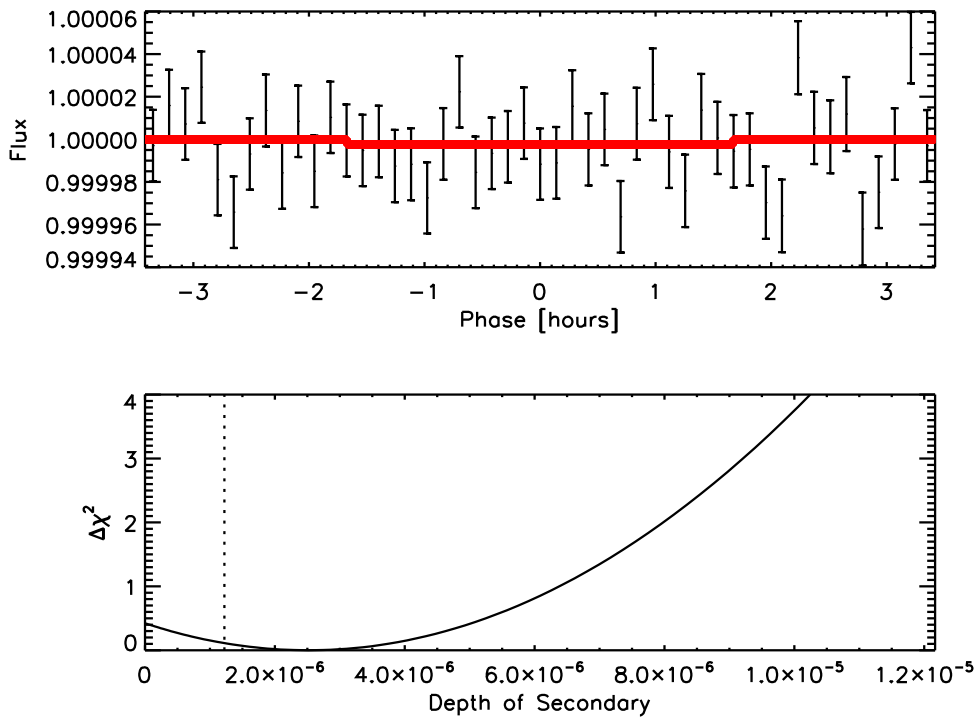} 
 \caption{The Kepler-19 light curve at a phase of 0.5. The best eclipse light curve is shown overplotted, but this solution is statistically indistinguishable from a flat line. The expected depth for planet with an albedo of one is shown by the dashed line.}
  \label{fig:secondary}
\end{center}
\end{figure}

To assess our sensitivity to the secondary eclipse, we fit a line to either side of each expected eclipse, and divided this line from the data during eclipse (centered on a phase of 0.5 and spanning 2.5 eclipse durations on either side), in a manner similar to the method we employed for data in transit. We then evaluated a model with epoch and duration set by the transit parameters given in Table \ref{tbl:params}, but with variable depth, from 0 to 20 ppm (considerably larger than the expected eclipse depth), and compared to the phase-folded light curve. We neglect the possibility that the transiting planet resides in an eccentric orbit, and thus that the secondary eclipse may not occur at a phase of 0.5. We find that our sensitivity is yet too low to detect the secondary eclipse at any phase. Values as high as 9 ppm are statistically indistinguishable from a depth of zero (furnishing a $\chi^{2}$ difference less than 9). A planet with a maximal Bond albedo of 1 would produce a decrement of 1.2 ppm, which is still considerably below detectability with \kepler. Figure \ref{fig:secondary} shows the $\chi^{2}$ improvement associated with adding an eclipse of variable depth at a phase of 0.5. The dotted line shows the expected depth for an extreme albedo of 1, which is indistinguishable from a flat line. 

\subsection{Composition of Kepler-19b}
While we cannot estimate the mean density of Kepler-19b without a measurement of its mass, we can still draw meaningful conclusions about its composition from the upper limit mass value, as was done by \cite{Fressin11} in the similar case of Kepler-10c. We first address whether we can rule out solid compositions at the highest density: a planet made of pure iron at a radius of 2.2 $R_{\oplus}$ would have a mass of 100 $M_{\oplus}$ \citep{Seager07}. However, such a high fraction of iron is unphysical, even with maximal collisional stripping during the planet's formation; a 2.2 $R_{\oplus}$ planet with the largest possible iron fraction would have a mass of 30 $M_{\oplus}$ \citep{Marcus10}. This maximum density is ruled out with 95\% confidence by the mass upper limit from radial velocities of 20.3 $M_{\oplus}$. A planet composed of pure silicate at the measured radius, however, would have a mass of 15 $M_{\oplus}$ \citep{Seager07}, which lies  within the allowable mass range for Kepler-19b. In contrast, a 2.2 $R_{\oplus}$ planet with a homogeneous composition of water ice would have a mass of 4.5 $M_{\oplus}$ \citep{Seager07}: mixtures of water ice and silicate in any fraction are therefore consistent with the measured mass upper limit. We consider also whether a substantial H/He envelope is possible for Kepler-19b. This scenario brackets the lower range of possible densities. \cite{Rogers11} present theoretical models for planets in the radius range of 2-6 $R_{\oplus}$ and the temperature range 500-1000 K (in which sample Kepler-19b, with radius of 2.2 $R_{\oplus}$ and temperature near 700 K, is included), given different formation histories. If Kepler-19b formed by a nucleated core-accretion scenario beyond the snow line, a core of ice and rock surrounded by an H/He envelope would not be tenable even at the cooler equilibrium temperature of 500 K: such an envelope would have been lost in a timescale of $<$1 Gyr. An outgassed hydrogen envelope, by comparison, is a plausible scenario over a timescale greater than 1 Gyr, although the mass fraction of such an envelope would be less than 0.01 the mass of the planet \citep{Rogers11}.

\subsection{Future Prospects}
In light of the dynamical study of Section 6.1, which combined transit time variations and radial velocities to characterize a perturber in the Kepler-19 system, we may ask whether we can expect to measure the precise orbital parameters of planet 19c. The two main issues causing ambiguity among the perturber scenarios are: (i) the transit variations are consistent with a smooth sinusoid down to the noise level, with no additional hints of the perturber's identity such as ``chopping'' \citep{Holman10} on the conjunction timescale, (ii) the radial velocities, while essential to ruling out massive, non-planetary perturbers, simply do not have the precision sufficient to distinguish between the various planetary scenarios. The former issue might be resolved with much more data. The \kepler\ collaboration intends to keep Kepler-19 on its short cadence mode (demonstrated here to result in superior timing accuracy) for the remainder of the mission. Furthermore, in numerical simulations, chopping signals can wax and wane over secular cycles, so a detection of this effect is possible in future quarters. The latter issue requires more radial velocity data to resolve, which could allow us to further address the degeneracy of interpretation of the TTV signal, or could point to even more planets that are not clearly detected by transit timing variations. 

We would like to thank the Spitzer team at IPAC and in particular Nancy Silbermann for scheduling the Spitzer observations of this program. This work is based on observations made with the Spitzer Space Telescope, which is operated by the Jet Propulsion Laboratory, California Institute of Technology under a contract with NASA. Support for this work was provided by NASA through an award issued by JPL/Caltech. This work is also based on observations made with Kepler, which was competitively selected as the tenth Discovery mission. Funding for this mission is provided by NASA's Science Mission Directorate. The authors would like to thank the many people who generously gave so much their time to make this Mission a success. Some of the data presented herein were obtained at the W.M. Keck Observatory, which is operated as a scientific partnership among the California Institute of Technology, the University of California and the National Aeronautics and Space Administration. The Observatory was made possible by the generous financial support of the W.M. Keck Foundation. 


\begin{deluxetable}{cccccc}
\tablenum{2}
\tablecaption{Transit Times for Kepler-19b From Q0-Q8}
\label{tbl:times}
\tablewidth{0pt}
\tablehead{
\colhead{Transit Number} & \colhead{Transit Time}    & \colhead{Predicted Time}  & \colhead{O-C}   & \colhead{$-1\sigma$}     & \colhead{$+1\sigma$}  \\
\colhead{ } & \colhead{[BJD-2450000]} & \colhead{(from linear ephemeris)} & \colhead{[min]} & \colhead{[min]} & \colhead{[min]} \\}
\startdata
1 & 4959.70744 & 4959.70605 & 2.0 & 2.6 & 2.7 \\
2 & 4968.98895 & 4968.99305 & -5.9 & 3.4 & 5.6 \\
3 & 4978.27949 & 4978.28004 & -0.8 & 3.6 & 3.2 \\
4 & 4987.56280 & 4987.56704 & -6.1 & 3.3 & 3.2 \\
5 & 4996.85396 & 4996.85403 & -0.10 & 3.8 & 3.5 \\
6 & 5006.13936 & 5006.14102 & -2.4 & 3.4 & 4.2 \\
8 & 5024.71126 & 5024.71501 & -5.4 & 3.2 & 3.3 \\
9 & 5033.99971 & 5034.00201 & -3.3 & 4.0 & 3.9 \\
10 & 5043.28879 & 5043.28900 & -0.30 & 3.2 & 3.5 \\
11 & 5052.57509 & 5052.57599 & -1.3 & 3.9 & 4.0 \\
12 & 5061.85993 & 5061.86299 & -4.4 & 3.2 & 3.3 \\
13 & 5071.14832 & 5071.14998 & -2.4 & 3.5 & 3.4 \\
14 & 5080.43524 & 5080.43698 & -2.5 & 3.3 & 4.5 \\
15 & 5089.72744 & 5089.72397 & 5.0 & 3.2 & 2.9 \\
16 & 5099.00860 & 5099.01096 & -3.4 & 2.2 & 1.7 \\
17 & 5108.29463 & 5108.29796 & -4.8 & 1.7 & 1.5 \\
18 & 5117.58342 & 5117.58495 & -2.2 & 2.0 & 1.7 \\
19 & 5126.87077 & 5126.87195 & -1.7 & 2.8 & 2.0 \\
20 & 5136.15957 & 5136.15894 & 0.9 & 4.1 & 1.5 \\
21 & 5145.44635 & 5145.44593 & 0.6 & 1.8 & 2.0 \\
23 & 5164.02277 & 5164.01992 & 4.1 & 2.0 & 0.9 \\
24 & 5173.30976 & 5173.30692 & 4.1 & 1.5 & 1.3 \\
26 & 5201.17269 & 5201.16790 & 6.9 & 1.5 & 1.3 \\
27 & 5210.45913 & 5210.45489 & 6.1 & 3.9 & 3.1 \\
28 & 5219.74591 & 5219.74189 & 5.8 & 1.4 & 1.4 \\
29 & 5229.03319 & 5229.02888 & 6.2 & 1.6 & 2.3 \\
30 & 5238.31921 & 5238.31587 & 4.8 & 1.3 & 1.3 \\
31 & 5247.60551 & 5247.60287 & 3.8 & 1.4 & 1.4 \\
32 & 5256.89243 & 5256.88986 & 3.7 & 1.5 & 1.8 \\
33 & 5266.18033 & 5266.17686 & 5.0 & 1.5 & 2.0 \\
34 & 5284.75126 & 5284.75084 & 0.6 & 2.6 & 1.6 \\
36 & 5294.03582 & 5294.03784 & -2.9 & 1.0 & 1.9 \\
37 & 5303.32205 & 5303.32483 & -4.0 & 1.7 & 2.2 \\
38 & 5312.61016 & 5312.61183 & -2.4 & 1.6 & 1.7 \\
39 & 5321.89569 & 5321.89882 & -4.5 & 2.0 & 1.3 \\
40 & 5331.18179 & 5331.18581 & -5.8 & 1.5 & 1.3 \\
41 & 5340.46857 & 5340.47281 & -6.1 & 2.6 & 1.7 \\
42 & 5349.75577 & 5349.75980 & -5.8 & 8.7 & 1.6 \\
43 & 5359.04291 & 5359.04680 & -5.6 & 1.6 & 1.7 \\
44 & 5368.33171 & 5368.33379 & -3.0 & 1.5 & 2.3 \\
45 & 5377.61939 & 5377.62078 & -2.0 & 1.8 & 1.5 \\
46 & 5386.90368 & 5386.90778 & -5.9 & 2.0 & 1.8 \\
47 & 5396.19179 & 5396.19477 & -4.3 & 2.4 & 1.7 \\
48 & 5405.47968 & 5405.48177 & -3.0 & 2.4 & 1.5 \\
49 & 5414.76688 & 5414.76876 & -2.7 & 2.2 & 1.5 \\
50 & 5424.05381 & 5424.05575 & -2.8 & 3.2 & 2.4 \\
51 & 5433.34275 & 5433.34275 & 0.0 & 1.8 & 2.1 \\
52 & 5442.62821 & 5442.62974 & -2.2 & 3.2 & 3.2 \\
53 & 5451.91715 & 5451.91674 & 0.6 & 2.0 & 1.4 \\
54 & 5461.20241 & 5461.20373 & -1.9 & 1.7 & 3.8 \\
55 & 5470.49253 & 5470.49072 & 2.6 & 1.8 & 1.9 \\
56 & 5479.77723 & 5479.77772 & -0.7 & 1.3 & 1.4 \\
57 & 5489.06589 & 5489.06471 & 1.7 & 2.4 & 3.1 \\
58 & 5498.35615 & 5498.35171 & 6.4 & 1.8 & 2.1 \\
59 & 5507.64127 & 5507.63870 & 3.7 & 1.3 & 1.2 \\
60 & 5516.92944 & 5516.92569 & 5.4 & 1.5 & 1.8 \\
61 & 5526.21866 & 5526.21269 & 8.6 & 2.1 & 1.8 \\
62 & 5535.49975 & 5535.49968 & 0.1 & 1.4 & 1.7 \\
63 & 5544.78855 & 5544.78668 & 2.7 & 1.3 & 1.2 \\
64 & 5572.64863 & 5572.64766 & 1.4 & 1.7 & 1.3 \\
67 & 5581.93437 & 5581.93465 & -0.4 & 1.4 & 1.8 \\
68 & 5591.22095 & 5591.22165 & -1.0 & 1.2 & 1.2 \\
69 & 5600.50788 & 5600.50864 & -1.1 & 1.3 & 1.4 \\
70 & 5609.79355 & 5609.79563 & -3.0 & 1.3 & 1.6 \\
71 & 5619.07832 & 5619.08263 & -6.2 & 1.4 & 1.7 \\
72 & 5628.36705 & 5628.36962 & -3.7 & 1.2 & 1.2 \\
\enddata
\tablecomments{Transits with numbers $<16$ were gathered at long cadence (with an exposure time of 29.5 minutes), while transits with numbers $\ge16$ were gathered at short cadence (with an exposure time of 58.8 seconds).}
\end{deluxetable}

\end{document}